%% file: quantized_inverse_design.tex
\documentclass[journal=jacsph,manuscript=article]{achemso}

\usepackage[version=3]{mhchem} %
\input{Templates/math_commands}

\graphicspath{quantized_inverse_design}

\usepackage{booktabs}
\usepackage{pifont}
\usepackage{xcolor}

\usepackage{caption}
\usepackage{subcaption}
\usepackage{easy-todo}

\usepackage{hyperref}
\hypersetup{
  colorlinks,
  citecolor=blue,
  linkcolor=black,
  urlcolor=black}
\usepackage[capitalise]{cleveref}
\crefname{todos}{todo}{todos}
\Crefname{todos}{Todo}{Todos}
\usepackage{siunitx}
\usepackage{adjustbox}
\usepackage[ruled,vlined]{algorithm2e}
\usepackage{amssymb}
\usepackage{soul}
\usepackage{ifthen}
\usepackage{mdframed}
\usepackage{etoolbox}
\usepackage{mathtools}
\usepackage{tcolorbox}

\SetKwComment{Comment}{// }{}
\crefname{algocf}{Alg.}{algorithms}
\Crefname{algocf}{Alg.}{Algorithms}

\definecolor{lightreda}{rgb}{1,0.8,0.8}
\newboolean{myflag}
\setboolean{myflag}{false}

\NewDocumentCommand{\highlight}{m}{%
  \ifbool{myflag}{%
    \begingroup
    \sethlcolor{lightreda}%
    \soulregister\cite{7}%
    \soulregister\citet{7}%
    \soulregister\footnote{7}%
    \soulregister\ref{7}%
    \soulregister\pageref{7}%
    \soulregister\label{7}%
    \soulregister\cref{7}%
    \hl{#1}%
    \endgroup
  }{%
    #1%
  }%
}

\newenvironment{heqn}{%
  \ifbool{myflag}{%
    \begin{mdframed}[backgroundcolor=lightreda,
                     hidealllines=true,
                     innerleftmargin=0pt,
                     innerrightmargin=0pt,
                     innertopmargin=-7pt,
                     innerbottommargin=-7pt]%
  }{%
    \relax
  }%
}{%
  \ifbool{myflag}{%
    \end{mdframed}%
  }{%
    \relax
  }%
}

\author{Frederik Schubert}
\email{schubert@tnt.uni-hannover.de}
\altaffiliation{Equal Contribution}
\author{Yannik Mahlau}
\email{mahlau@tnt.uni-hannover.de}
\altaffiliation{Equal Contribution}
\author{Konrad Bethmann}
\author{Fabian Hartmann}
\affiliation[Leibniz University]
{Institute for Information Processing, Leibniz University, Hannover, Germany}
\author{Reinhard Caspary}
\affiliation[PhoenixD]
{PhoenixD, Leibniz University, Hannover, Germany}
\author{Marco Munderloh}
\affiliation[Leibniz University]
{Institute for Information Processing, Leibniz University, Hannover, Germany}
\author{Jörn Ostermann}
\author{Bodo Rosenhahn}
\affiliation[Leibniz University]
{Institute for Information Processing, Leibniz University, Hannover, Germany}

\title[Quantized Inverse Design]{Quantized Inverse Design for Photonic Integrated Circuits}

\abbreviations{FDTD, AD}

\begin{document}

\input{quantized_inverse_design/sections/00_abstract.tex}

\newpage

\input{quantized_inverse_design/sections/01_introduction.tex}

\input{quantized_inverse_design/sections/02_methods.tex}

\input{quantized_inverse_design/sections/03_results.tex}

\input{quantized_inverse_design/sections/04_conclusion.tex}

\begin{acknowledgement}

This work was supported by the Federal Ministry of Education and Research (BMBF), Germany under the AI service center KISSKI (grant no. 01IS22093C), the Lower Saxony Ministry of Science and Culture (MWK) through the zukunft.niedersachsen program of the Volkswagen Foundation and the Deutsche Forschungsgemeinschaft (DFG) under Germany’s Excellence Strategy within the Cluster of Excellence PhoenixD (EXC 2122) and  (RO 2497/17-1). Additionally, this was funded by the Deutsche Forschungsgemeinschaft (DFG, German Research Foundation) – 444745111 and 517733257.

\end{acknowledgement}

\begin{suppinfo}

\input{quantized_inverse_design/sections/05_appendix.tex}

\end{suppinfo}

\bibliography{PhoenixD2}

\end{document}

%% file: Templates/math_commands.tex
\usepackage{amsmath,amsfonts,bm}

\def\eqref#1{equation~\ref{#1}}

\def\1{\bm{1}}

\def\vs{{\bm{s}}}

\DeclareMathAlphabet{\mathsfit}{\encodingdefault}{\sfdefault}{m}{sl}
\SetMathAlphabet{\mathsfit}{bold}{\encodingdefault}{\sfdefault}{bx}{n}
\newcommand{\tens}[1]{\bm{\mathsfit{#1}}}

\def\tE{{\tens{E}}}

\def\tH{{\tens{H}}}

\def\tJ{{\tens{J}}}

\def\tP{{\tens{P}}}

\DeclareMathOperator*{\argmin}{arg\,min}

%% file: quantized_inverse_design/sections/00_abstract.tex
\begin{abstract}
The inverse design of photonic integrated circuits (PICs) presents distinctive computational challenges, including their large memory requirements.
Advancements in the two-photon polymerization (2PP) fabrication process introduce additional complexity, necessitating the development of more flexible optimization algorithms to enable the creation of multi-material 3D structures with unique properties.
\highlight{
This paper presents a memory efficient reverse-mode automatic differentiation framework for finite-difference time-domain (FDTD) simulations that is able to handle complex constraints arising from novel fabrication methods.
}
Our method is based on straight-through gradient estimation that enables non-differentiable shape parametrizations.
We demonstrate the effectiveness of our approach by creating increasingly complex structures to solve the coupling problem in PICs.
The results highlight the potential of our method for future PIC design and practical applications.
\end{abstract}

%% file: quantized_inverse_design/sections/01_introduction.tex
\section{Introduction}

The availability of commercial two-photon polymerization (2PP) printers \cite{janduraPolymerBasedDevices2017,gonzalez-hernandezMicroOptics3D2023} has led to an increased interest in the production and optimization of photonic integrated circuits (PICs) based on polymers.
In comparison to silicon-based PICs, polymer-based PICs offer advantages in terms of prototyping and material costs \cite{zhangPolymerBasedHybridIntegratedPhotonic2013}.
However, there are also disadvantages such as the high attenuation of the polymers compared to silicon-based components \cite{swatowskiPOLYMERWAVEGUIDEMANUFACTURING2017}.
A primary constraint is their small refractive index contrast, which necessitates the use of relatively large devices in comparison to the wavelength \cite{guenther_2017}.
Furthermore, light is coupled in and out using butt coupling or gratings\cite{marchettiCouplingStrategiesSilicon2019,sonHighefficiencyBroadbandLight2018} which have a large spatial footprint and further complicate the optical circuits.
This limitation affects the complexity of the inverse design process using techniques such as the finite-differences time-domain (FDTD) method \cite{kaneyeeNumericalSolutionInitial1966} since the interaction of light and material must be modeled at sufficiently high resolution over the simulated volume.
Developments such as the multi-material 2PP process can alleviate this problem of low refractive contrast by enabling more complex ways of manipulating the incoming light \cite{yangMultimaterialMultiphoton3D2021,gonzalez-hernandezSingleStep3D2023}.
Moreover, it also allows for the integration and optimization of active components due to the possibility of incorporating quantum dots into the polymer matrix \cite{yuQuantumDotsFacilitate2023}.
Nevertheless, the inverse design of such structures presents a challenge.
Other proposed methods such as the adjoint method \cite{AnsysLumericalFDTD,meep} are not directly applicable as they require the analytical derivation of operators for more complicated design spaces.
The advancements in highly parallel hardware such as GPUs and the respective efficient software \cite{agrawalTensorFlowEagerMultiStage2019,paszkePyTorchImperativeStyle2019,jax2018github} have made reverse-mode automatic differentiation (AD) a viable option, 
\highlight{
despite its limitations in terms of memory requirements when using a naive implementation.
This paper proposes a memory efficient framework that combines the flexibility and scalability of automatic differentiation with the generality of the FDTD method.
Our method employs an efficient implementation of reverse-mode AD using the time-reversibility of Maxwell's equations.
The method scales from a single machine to whole GPU clusters and is able to optimize a wide range of objective functions, design parameter spaces and constraints.
The flexibility of our method results from the straight-through gradient estimator which has been successfully applied to the 2D optimization problem of silicon PICs \cite{schubertInverseDesignPhotonic2022}.
We demonstrate the versatility of our method by optimizing the design of several polymer-based vertical couplers which are essential for the implementation and fabrication of PICs.
}

\begin{figure}[t]
    \centering
    \includegraphics[width=0.9\linewidth]{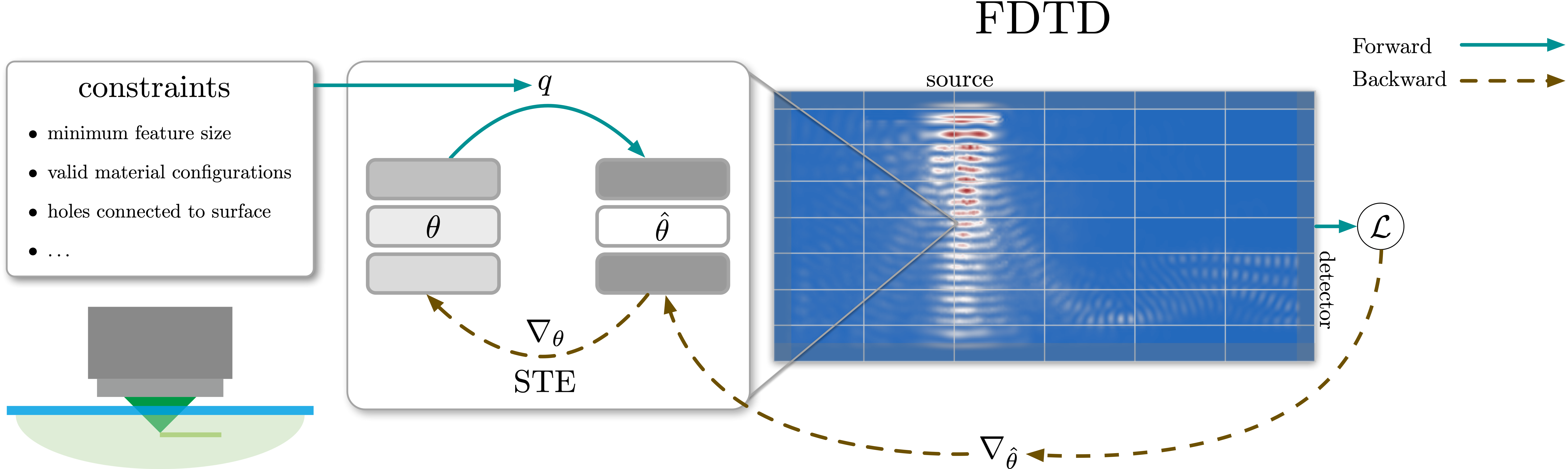}
    \caption{Overview of the proposed method. The fabrication constraints are encoded in the quantization mapping $q$ which transforms the latent parameters $\theta$ into the discrete realization $\hat{\theta}$.
    The fields are propagated via the FDTD simulation and the gradients of the objective function $\mathcal{L}$ are backpropagated through the simulation. 
    Finally, the gradients are transformed via the straight-through estimator (STE) to update the latent parameters.}
    \label{fig:overview}
\end{figure}

\highlight{
In summary, our main \textbf{contribution} is a memory efficient implementation of reverse-mode automatic differentiation.
This enables the creation of a general quantization-based framework to integrate arbitrary constraints in 3D into the inverse design process of large-scale photonic integrated circuits.
The framework scales from single GPUs to GPU clusters and is published as open source to spur further research in this area\footnote{The code is publicly available at \url{https://github.com/ymahlau/fdtdx}}.
We demonstrate the capabilities of our approach by the inverse design of several coupling devices that range from a simple silicon coupler to a multi-material polymer coupler with complex fabrication constraints.
Our results indicate the potential of multi-material 2PP for the fabrication of photonic integrated circuits.
}

%% file: quantized_inverse_design/sections/02_methods.tex
\section{Methods}

Our work is based on the finite-differences time-domain method, which we briefly introduce in the following section.
We also provide an overview of the inverse design problem for photonic integrated circuits including a discussion of the capabilities and limitations of previous approaches that use the adjoint method or automatic differentiation.
Finally, we introduce our method to integrate arbitrary constraints into the inverse design process.

\subsection{The Finite-Differences Time-Domain Method}

Our framework is based on the finite-differences time-domain (FDTD) \cite{kaneyeeNumericalSolutionInitial1966} method.
FDTD solves the classical electrodynamics description of Maxwell's equations \cite{maxwellVIIIDynamicalTheory1865} using leapfrog integration.
The electric and magnetic fields $\tE$ and $\tH$ are discretized and arranged in the staggered Yee Lattice.
Given an initial state, the fields are integrated in discrete time steps in an interleaved pattern.
The time step duration $\Delta t$ is determined according to the Courant Levy stability conditions \cite{courantUberPartiellenDifferenzengleichungen1928}, which ensure that information does not propagate faster than the speed of light.
The update equations for the fields are given by Maxwell's equations:
\begin{heqn}
\begin{align}
    \nabla \times \tE &= -\mu_0\tens{\mu}_r \frac{\partial \tH}{\partial t} \\
    \nabla \times \tH &= \epsilon_0\tens{\epsilon}_r \frac{\partial \tE}{\partial t} + \tJ
\end{align}
\end{heqn}
\highlight{
where $\tE$ and $\tH$ are discretized in space and time.
}
The fields in the simulation are modulated by permittivity and permeability tensors, $\tens{\epsilon}_r$ and $\tens{\mu}_r$, which are also discretized onto the Yee grid.
When the fields are excited by an electric source $\tJ$, the energy follows the dynamics of Maxwell' equations and travels in waves throughout the simulation.
To simulate an infinitely large space, we apply convolutional Perfectly Matched Layers (CPMLs) at the boundaries of our simulation \cite{rodenConvolutionPMLCPML2000}.
Accurately modeling the fields for a given source wavelength $\lambda_0$ sets bounds on the spatial resolution $\Delta \vs$ that should be approximately $\lambda_0 / 10$ \cite{tafloveComputationalElectrodynamicsFiniteDifference1998} but has to be adapted to the specific problem.
For metrics that combine the electric and magnetic field, such as the poynting flux $\tP = \tE \times \tH$, the fields from the leapfrog integration have to be synchronized in time and interpolated on the same spatial grid coordinates from the Yee lattice.

\subsection{Inverse Design for Photonic Integrated Circuits}

The inverse design of photonic integrated circuits (PICs) can generally be formulated as a constrained \emph{topology optimization} problem.
In the context of polymer-based PICs, the design space is the discrete distribution of polymer in a fixed volume to maximize a given objective function $\mathcal{L}$.
Many methods are gradient-based, i.e. they use the local sensitivity of the objective with respect to the design parameters $\theta$ to guide their optimization.
A common design parametrization in photonic inverse design is the permittivity tensor of the design region from which the manufacturable device can be derived.
The predominant method for gradient-based optimization in the context of FDTD simulations is the adjoint variable method \cite{AnsysLumericalFDTD,meep,EMoptDocumentationEMopt} due to its efficiency and scalability.
This, however, restricts the design space to smooth and differentiable functions such as level-set methods \cite{lalau-keralyAdjointShapeOptimization2013,christiansenInverseDesignPhotonics2021} where the gradients of the adjoint fields with respect to the parameters can be derived analytically \cite{guoqiangshenAdjointSensitivityTechnique2003,bakrAdjointVariableMethod2003,nikolovaSensitivityAnalysisFDTD2004,nikolovaSensitivityAnalysisScattering2006,swillamAdjointSensitivityAnalysis2007,yasudaDesignMethodSpatiotemporal2019}.
The recently proposed time-reversal direct differentiation method \cite{tangTimeReversalDifferentiation2023} suffers from the same limitations.
A method to circumvent this constraint is the combination of the adjoint method with automatic differentiation \cite{luceMergingAutomaticDifferentiation2023,hootenAutomaticDifferentiationAccelerated2023}.
This releases the user from deriving the analytical gradients and allows for a more flexible design space.
However, all of the above methods are limited to continuous design spaces.
Some works use discrete optimization methods such as \citet{ballewConstrainingContinuousTopology2023} but introduce additional complexity via an additional inner optimization problem and are not as efficient as first-order gradient-based methods.

In this work, we propose a framework to efficiently design devices with arbitrary constraints.
Following \citet{schubertInverseDesignPhotonic2022a} and \citet{paoliniPhotonicawareNeuralNetworks2022}, we apply the straight-through gradient estimator (STE) \cite{bengioEstimatingPropagatingGradients2013,yinUnderstandingStraightEstimatorTraining2019} to implement our constrained optimization problem
\begin{equation}
\frac{\partial \mathcal{L}}{\partial \theta} = \frac{\partial \mathcal{L}}{\partial \hat{\theta}} \frac{\partial \hat{\theta}}{\partial \theta} \approx \frac{\partial \mathcal{L}}{\partial \hat{\theta}}.
\label{eq:ste}
\end{equation}
The STE allows us to backpropagate gradients through non-differentiable functions $q$ by replacing the transformation of the gradient in the backward pass with the identity function.
\highlight{
This introduces a gradient error, but leads to approximately equivalent latent parameter movement when using adaptive learning rate optimizers \cite{Schoenbauer2024CustomGE} like Adam \cite{KingmaB14}.
}
The STE is particularly useful for the FDTD simulation, as it allows us to backpropagate gradients through the discrete design parameters $\hat{\theta} = q(\theta)$ of the simulation into the continous latent parameters $\theta$.
An overview of our framework is shown in \cref{fig:overview}.

\begin{algorithm}
\caption{\highlight{2D Quantization Mapping}}\label{alg:2d}
\KwIn{Latent parameters $\theta$, valid permittivities $\mathcal{P}$}
\KwOut{Quantized parameters $\hat{\theta}$}
\For{$(x,y) \in X \times Y$}{
$\hat{\theta}_{x, y} \gets \argmin\limits_{{p} \in \mathcal{P}} |\theta_{x, y} - p|$\;
}
\Return{$\hat{\theta}$}\;
\end{algorithm}

\highlight{
In this work, we use a single scalar $\theta_{x,y,z} \in \mathbb{R}$ as a latent variable for each voxel in the topology optimization region.
Using the quantization function $q$, the latent variables $\theta$ are mapped to quantized parameters $\hat{\theta}$.
Specifically, the condition $\hat{\theta}_{x,y,z} \in \mathcal{P}$ has to hold for all $(x, y, z) \in X \times Y \times Z$, where $\mathcal{P}$ is the set of valid permittivities.
For example, when designing single material PIC using the ma N 1400 Series polymer \cite{microresisttechnologyProcessingGuidelines}, valid permittivities would be $\mathcal{P} = \{1.0, 1.608\}$ representing the binary choice between air or material.
The quantization mapping $q$ to the closest valid permittivity is displayed in \cref{alg:2d}.
The more complex quantization constraints for 2.5D and 3D are described in the later sections.
}

\begin{algorithm}
\caption{\highlight{Inverse Design with STE}}\label{alg:optim}
\KwIn{valid permittivities $\mathcal{P}$, epoch count $n \in \mathbb{N}_{>0}$, learning rate schedule $\alpha$}
\KwIn{Quantization mapping $q$, Objective function $\mathcal{L}$}
\KwOut{Optimized parameters $\theta$}
\smallskip
\Comment{Random Initialization}
\For{$(x,y,z) \in X \times Y \times Z$}{
$\theta_{x,y,z} \gets $ random sample in the range $[\min_{p \in \mathcal{P}} p, \max_{p \in \mathcal{P}} p]$
}
\smallskip
\Comment{Optimization}
\For{$i \gets 0$ \KwTo $n - 1$}{
$\hat{\theta} \gets q(\theta)$\;
$\mathcal{L}(\hat{\theta}) \gets$  FDTD simulation using $\hat{\theta}$\;
$\frac{\partial \mathcal{L}}{\partial \hat{\theta}} \gets$ time-reversible automatic differentiation\;
$\theta \gets \theta + \alpha_i \frac{\partial \mathcal{L}}{\partial \hat{\theta}}$
}
\Return{$\theta$}\;
\end{algorithm}

\highlight{
In \cref{alg:optim}, our full optimization algorithm is shown.
Firstly, the latent parameters $\theta$ are initialized randomly between the minimum and maximum valid permittivity.
Then, an optimization loop performs gradient ascent for $n$ iterations.
In each iteration, the parameters $\theta$ are quantized and run through the FDTD simulation.
Using a memory efficient time-reversible implementation of automatic differentiation, we calculate the gradient $\frac{\partial \mathcal{L}}{\partial \hat{\theta}}$ of the objective function $\mathcal{L}$ with respect to the quantized parameters.
With the STE-approximation, we can use this gradient to update the latent parameters.
We determine the learning rates $\alpha_i$ using the Adam optimizer \cite{KingmaB14} with Nesterov momentum \cite{1370862715914709505} and cosine-decay learning rate schedule with linear warmup \cite{LoshchilovH17}.
}

\subsubsection{Memory Efficient Implementation}
\highlight{
Calculating the gradient through an FDTD simulation using direct differentiation can be very memory intensive.
A naive implementation requires the electric and magnetic field to be saved after every time step.
We circumvent this problem by using the time-reversible nature of Maxwell's equations \cite{sorrentinoTimeReversalFinite1993}.
With this technique it is possible to calculate an inverse time step of the FDTD simulation that transforms the electric and magnetic fields at time step $t+1$ back to the fields at time step $t$.
As a result, only the fields of two time steps need to be kept in memory at any time
}
\footnote{\highlight{The simulation calculations in forward and backward direction also produce intermediate tensors, which require additional memory. Furthermore, the cumulative gradients need to be kept in memory.}}.
\highlight{
Unfortunately, the Perfectly Matched Layers (PML) at the boundary are not time-reversible.
Therefore, it is necessary to save the electric and magnetic fields at the boundary between PML and simulation volume for every time step to prevent information loss.
Even with this downside, the time-reversible gradient computation is more memory efficient than the naive implementation, because saving the 2D-slices between PML and simulation volume requires much less memory than saving the full 3D simulation volume. 
Previously, this technique has been successfully demonstrated by \citet{tangTimeReversalDifferentiation2023}, but has never been integrated into a powerful automatic differentiation framework like JAX \cite{jax2018github}.
The flexibility of automatic differentiation and the memory efficiency of time-reversible gradient computation result in an user-friendly workflow on relatively small hardware.
}

\begin{figure}[!t]
    \centering
    \begin{subfigure}{0.45\linewidth}
        \centering
        \includegraphics[width=\linewidth]{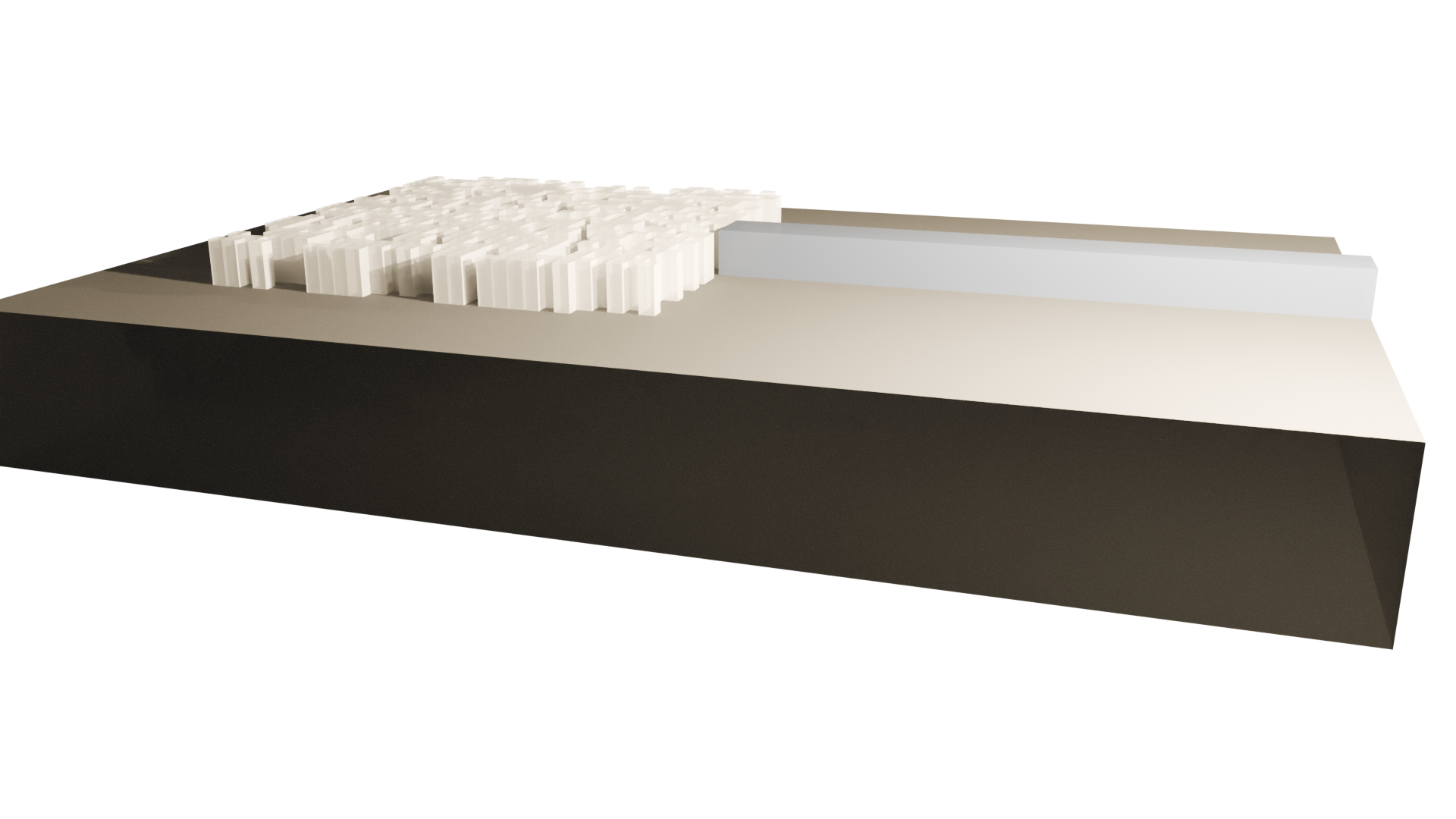}
        \caption{3D rendering of the silicon coupler on the silicon-oxide substrate with the single-mode waveguide shaded in grey.}
        \label{fig:silicon_coupler_setup}
    \end{subfigure}
    \hspace{1cm}
    \begin{subfigure}{0.45\linewidth}
        \centering
        \includegraphics[width=\linewidth]{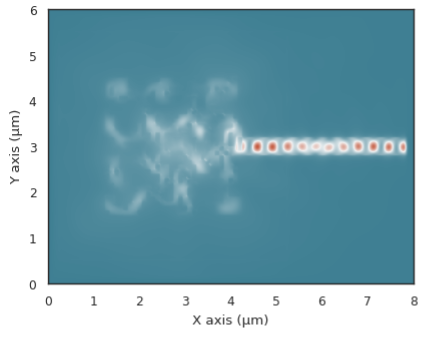}
        \caption{Snapshot of the energy distribution in the XY-plane of the optimized 2D silicon coupler. }
        \label{fig:silicon_coupler_structure}
    \end{subfigure}
    \caption{Reproduction of the silicon coupler from \citet{shenIntegratedMetamaterialsEfficient2014}. The structure consists of $30 \times 30$ cells of size \SI{100}{\nano\meter^2} that are either silicon or air. The device is optimized for a wavelength of \SI{1550}{\nano\meter} and a structure height of \SI{300}{\nano\meter}.}
    \label{fig:silicon_coupler}
\end{figure}

\subsubsection{Demonstration}

To demonstrate our framework in the domain of silicon photonics, we reproduce the coupling device from \citet{shenIntegratedMetamaterialsEfficient2014}.
The device consists of $30 \times 30$ square cells with a side length of \SI{100}{\nano\meter} that are either silicon or air.
The objective is the transfer efficiency from a free space planar wave into a single-mode silicon waveguide with a diameter of \SI{400}{\nano\meter}.
The device is optimized for a wavelength of \SI{1550}{\nano\meter} and a structure height of \SI{300}{\nano\meter}.
\highlight{
By starting a new optimization from random parameters, we match the reported results by \citet{shenIntegratedMetamaterialsEfficient2014} with a loss of $\SI{-2.7}{\deci\bel}$ compared to their $\SI{-3}{\deci\bel}$.
}
The resulting coupler is shown in \cref{fig:silicon_coupler}.

%% file: quantized_inverse_design/sections/03_results.tex
\begin{figure}[!t]
    \centering
    \includegraphics[width=\textwidth]{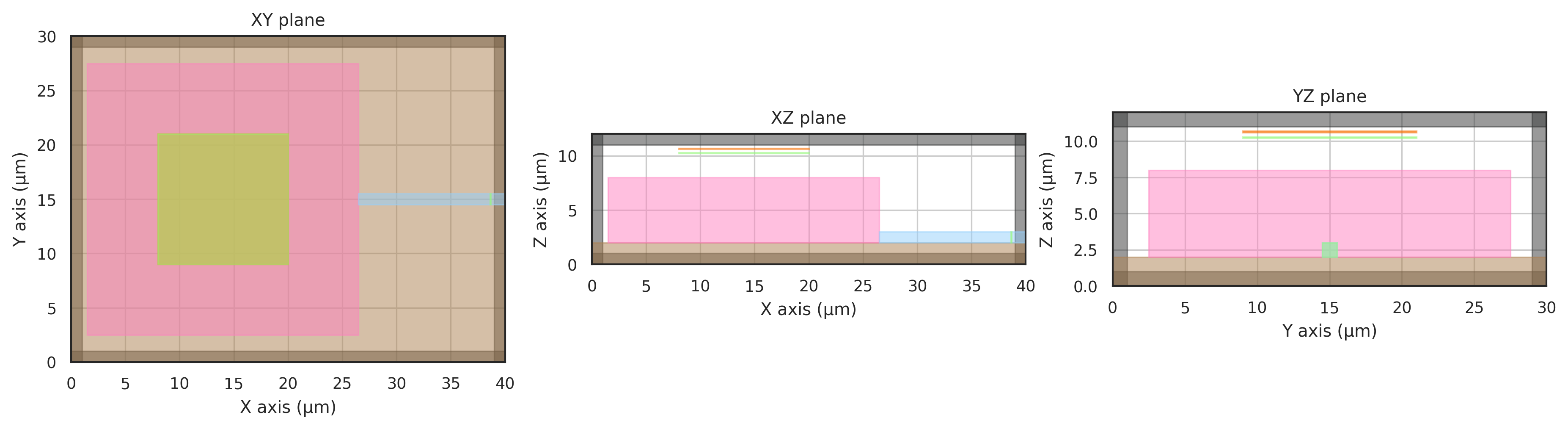}
    \caption{Setup of the vertical coupler with the free field source region (orange) above the device (pink). The device is connected to the output waveguide (blue) and is placed on top of a fused silica substrate (brown). Its efficiency is evaluated as the average poynting flux at the end of the waveguide (green) normalized by the incident poynting flux at the detector below the source. The reference plane on the input side (green) is placed at a small distance to the device to minimize the influence of reflections from the device during the optimization.}
    \label{fig:setup}
\end{figure}

\section{Results}
\label{sec:results}

To demonstrate the capabilities of our framework, we present three different examples of photonic integrated circuits (PICs).
The goal is to find a solution to the coupling problem, i.e. the efficient transfer of light from an external source into the photonic circuit.

\begin{figure}[!t]
    \centering
    \includegraphics[width=\textwidth]{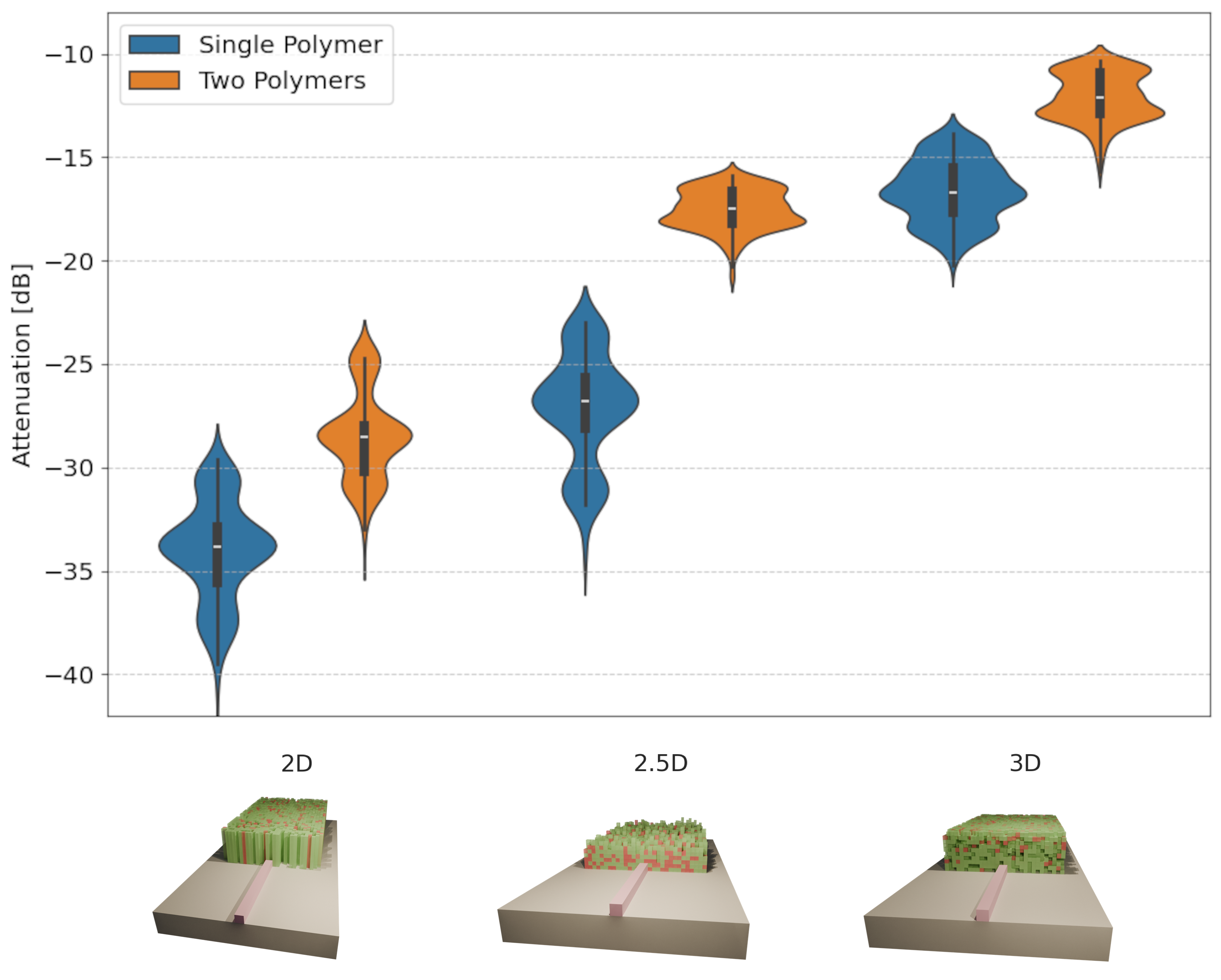}
    \caption{Comparison of coupler designs showing increasing efficiency benefits from a single polymer and two polymers for different design constraints. The median is computed over 50 evaluations per design and is due to the angle of the incoming light as well as the transverse displacement of the light source. The cavities in the 3D structures are always connected to the outer volume which allows for the removal of the unpolymerized photoresist.}
    \label{fig:teaser}
\end{figure}

They range from a simple single material vertical coupler with 2D and 2.5D structures to a full 3D two-polymer vertical coupler.
\highlight{
We refer to 2.5D as a 2D pixel grid, where each pixel may be extruded to a different height in the third dimension\cite{leeApplicationComputationalGeometry1995, siuShapeCADAccessible3D2019, siuAccessibleCADWorkflow2018}.
}
The general setup of the coupler is shown in \cref{fig:setup}.
We simulate a continuous source with a gaussian intensity distribution of \SI{4}{\micro\meter} radius and a wavelength of \SI{1.55}{\micro\meter} at a grid resolution of \SI{100}{\nano\meter}.
The source is placed at the region indicated in \cref{fig:setup} (orange rectangle in the XY plane) with random lateral offsets of \SI{2}{\micro\meter}.
\highlight{
Our setup consists of a 3D simulation volume of $\SI{40}{\micro\meter} \times \SI{30}{\micro\meter} \times \SI{12}{\micro\meter}$, which results in 14.4M grid cells.
}
Note, that the device is larger than the silicon coupler by a factor of about 10 due to the low refractive index contrast of the polymer.
Device and waveguide are placed on a fused silica substrate with a refractive index of $1.45$.
The objective function is the average poynting flux at the end of the output waveguide normalized by the incoming poynting flux at the detector below the source.
The output waveguide is rectangular with a width and height of \SI{1}{\micro\meter}.

\highlight{
The simulation is run for 250 femtoseconds with a Courant factor of $0.99$ which results in 1311 FDTD iterations.
We average the gradients over 3 simulations with random lateral source offsets to simulate calibration errors in our optimization process.
Each optimization iteration takes approximately $\SI{110}{\second}$ on a single NVIDIA A100.
}

\highlight{
For these optimizations, the design parameters $\theta$ of the coupler are the distributions of the polymer in the volume of the device.
The device is constrained by the minimum  feature size in the XY plane of $\SI{500}{\nano\meter}$ and either the possibility of using the full 3D design space or just the 2D and 2.5D solutions that are common in the literature.
For 2.5D structures, we impose a constraint that prevents any overhanging elements.
While the 3D design space offers more flexibility, it can potentially generate impractical designs with floating materials or cavities that would trap unpolymerized material.
To ensure feasibility in 3D designs, we incorporate these physical constraints directly into the quantization mapping.
}

\subsubsection{Single Material Vertical Coupler in 2D and 2.5D}

\begin{figure}[!t]
    \centering
    \includegraphics[width=0.9\textwidth]{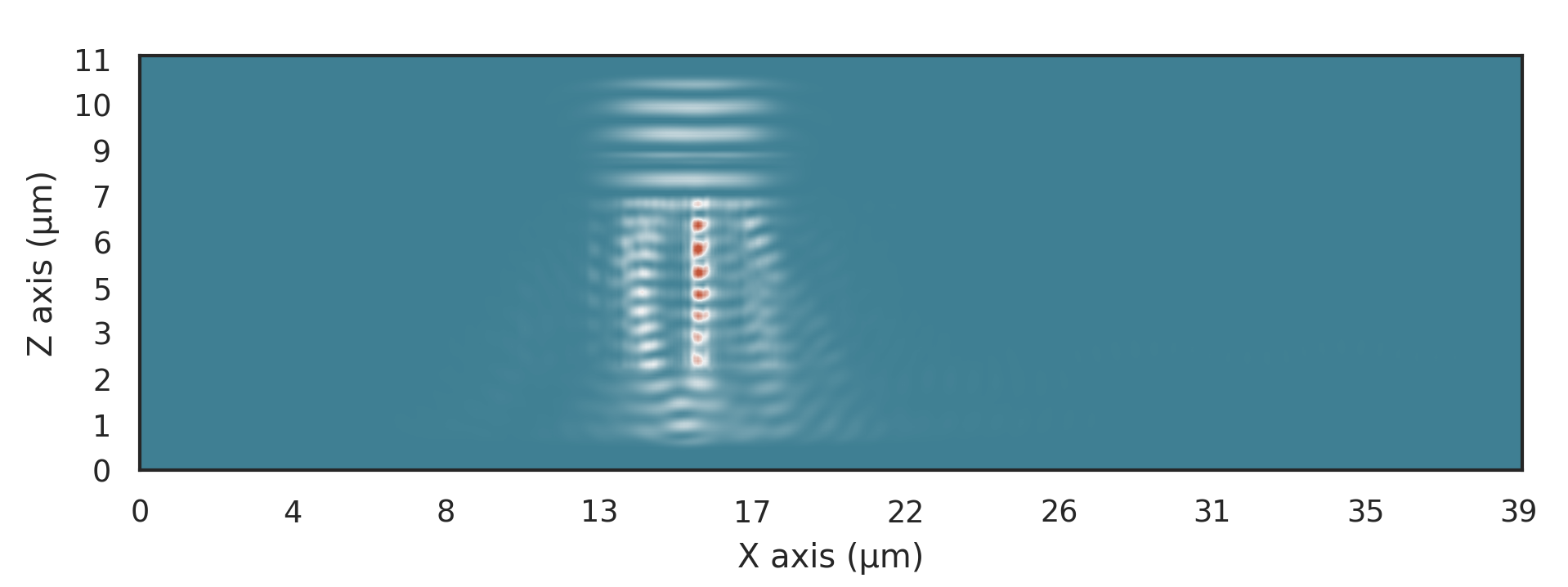}
    \caption{Energy distribution of the optimized 2D single material vertical coupler with a height of \SI{5}{\micro\meter} and a final attenuation of \SI{-34}{\deci\bel}. Almost no light is coupled into the waveguide to the right due to the low refractive index contrast.}
    \label{fig:2D_single_material_design}
\end{figure}

The efficiency of a vertical coupler depends on its ability to manipulate the incoming light via changes in the refractive index.
In this experiment, the number of ways for the coupler to interact with the light is limited by the single material design using the ma N 1400 Series polymer \cite{microresisttechnologyProcessingGuidelines}.
The results in \cref{fig:2D_single_material_design} show that the 2D distribution of polymer does not provide the required degrees of freedom to guide the light into the waveguide.
With an attenuation of \SI{-34}{\deci\bel}, almost all the light is lost to the substrate below the device.

\begin{figure}[!t]
\centering
\includegraphics[width=0.9\textwidth]{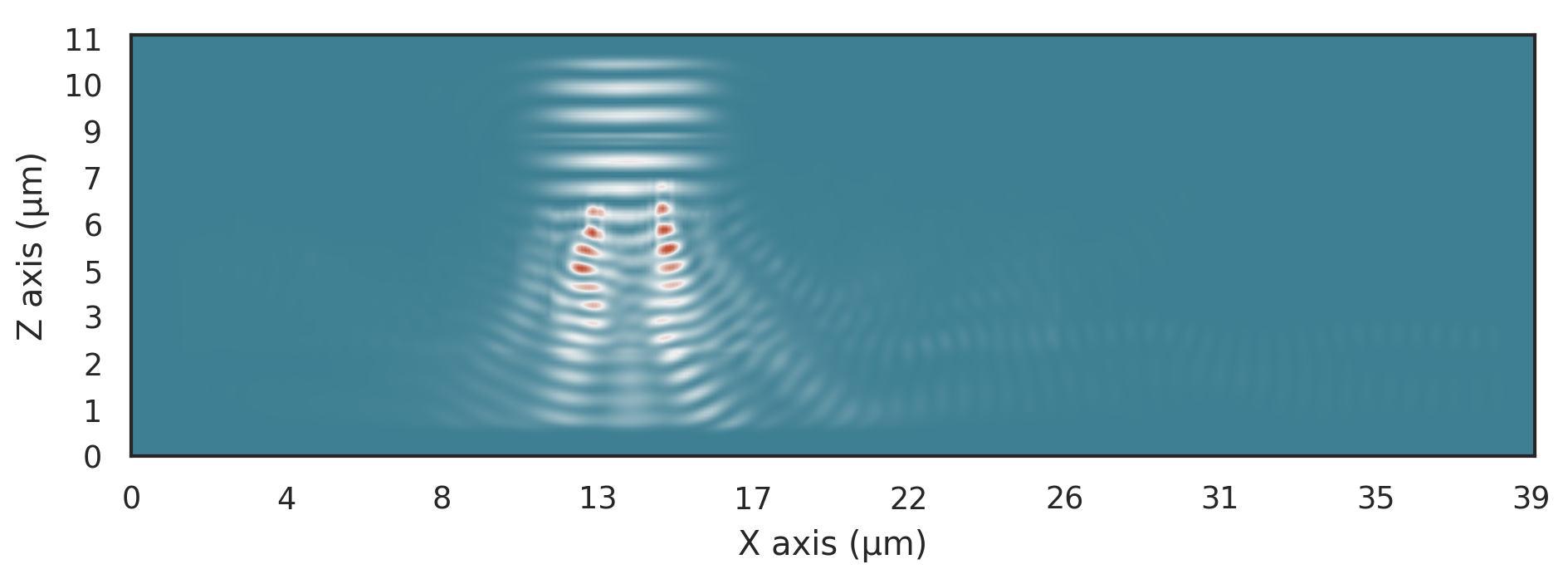}
\caption{Energy distribution of the optimized 2.5D single material coupler. 
The gaussian plane source above the device is visibly coupled into the single-mode waveguide to the right.
However, the efficiency of \SI{-29}{\deci\bel} remains low with significant energy loss to the substrate below and the surrounding volume.}
\label{fig:2.5D_single_energy}
\end{figure}

To improve the efficiency and to demonstrate the versatility of our quantization method, we extend our design space to 2.5D.
The results in \cref{fig:2.5D_single_energy} demonstrate that the efficiency of the 2.5D single material coupler is still insufficient, with almost all the energy being lost to the substrate and the volume around the device. 
Thus, this experiment highlights the need for more advanced designs with higher degrees of freedom to achieve efficiencies comparable to those in silicon PICs.

\subsubsection{Two-Polymer Vertical Coupler}

In this experiment, we demonstrate the inverse design of multiple polymers in the same device.
For this process, the polymers are chosen such that the refractive index between them is as high as possible.
We simulate the SZ 2080 polymer with a refractive index of $\approx 1.5$\footnote{The exact refractive index at \SI{1550}{\nano\meter} of the resist depends on the degree of polymerization which itself depends on the exact printing parameters \cite{gonzalez-hernandezMicroOptics3D2023}. Also see our brief section on 2PP printing in the supporting information.} and the ma N 1400 Series polymer with a refractive index of $1.608$\footnote{The Cauchy equation for the ma N 1400 Series is $n(\lambda) = 10^{-3} n_0 + 10^2 \frac{n_1}{\lambda^2} + 10^7 \frac{n_2}{\lambda^4}$ \cite{microresisttechnologyProcessingGuidelines}.
}.

\begin{figure}[!t]
    \centering
    \includegraphics[width=0.9\textwidth]{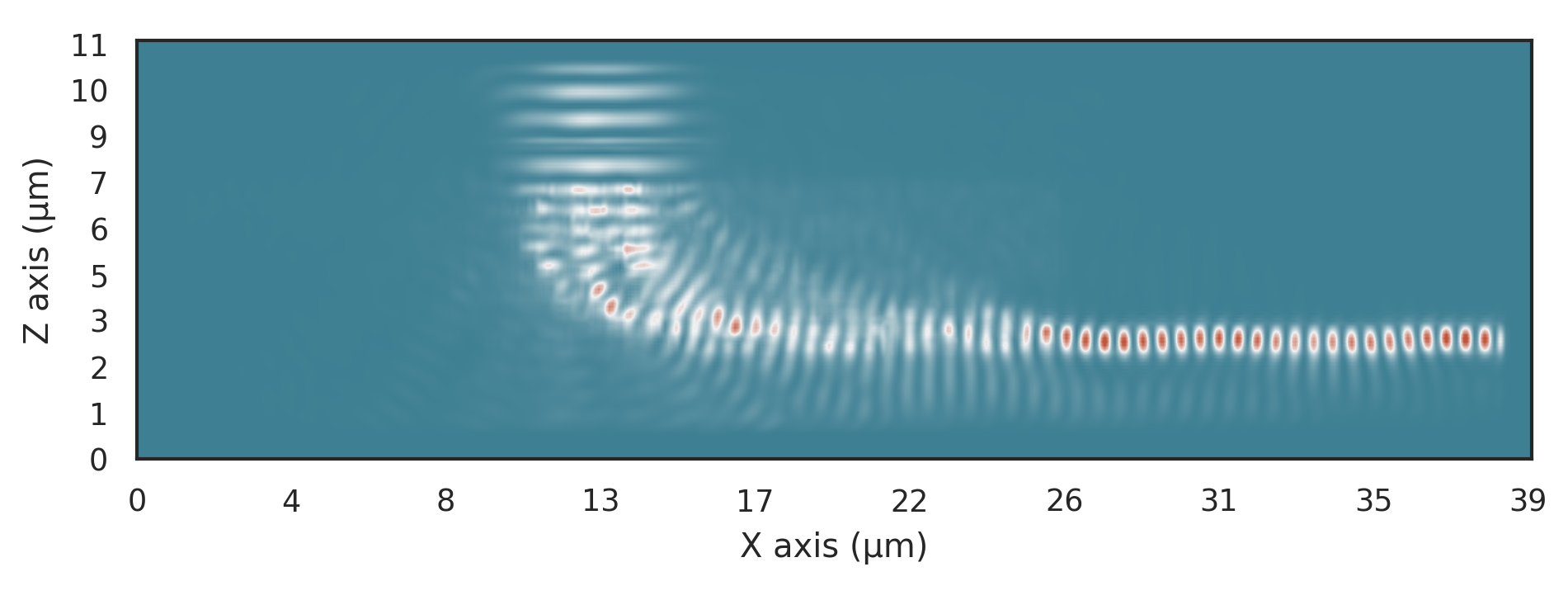}
    \caption{Energy distribution of the 3D two-polymer vertical coupler with a voxel height of \SI{500}{\nano\meter} and a final attenuation of \SI{-9}{\deci\bel}.}
    \label{fig:multi_material_designs}
\end{figure}

While the results for the 2D case are similar to that of the single material coupler, as indicated in \cref{fig:teaser}, the 2.5D design achieves a coupling efficiency of \SI{-17}{\deci\bel} or around 2\%.
However, to increase the theoretical efficiency even further, we use the full 3D design space of the two polymer device.
The results in \cref{fig:multi_material_designs} demonstrate the effectiveness of this design space by coupling over 12\% of the incoming light into the single-mode waveguide.
Much of the lost energy is due to the relatively large feature size of \SI{500}{\nano\meter}.
However, we wanted to present a device that can, in theory, be fabricated with the current state of the art in 2PP printing.

%% file: quantized_inverse_design/sections/04_conclusion.tex
\section{Conclusion}
\highlight{
In this work, we presented a memory efficient implementation of reverse-mode automatic differentiation, which enabled the integration of arbitrary constraints into the inverse design process of photonic integrated circuits.
}
Our approach is based on the quantization of the design parameters and the application of the straight-through gradient estimator to backpropagate gradients through the FDTD simulation.
We demonstrated the capabilities of our method by optimizing the design of several polymer-based vertical couplers.
These devices are core components of photonic integrated circuits and our results indicate the potential of multi-polymer 2PP for their fabrication.
Our method accelerates the development of polymer-based PICs due to its generality and usability.

%% file: quantized_inverse_design/sections/05_appendix.tex
\section{Validation of the FDTD Solver}
\label{sec:validation}

\begin{figure}[!t]
    \centering
    \includegraphics[width=0.9\textwidth]{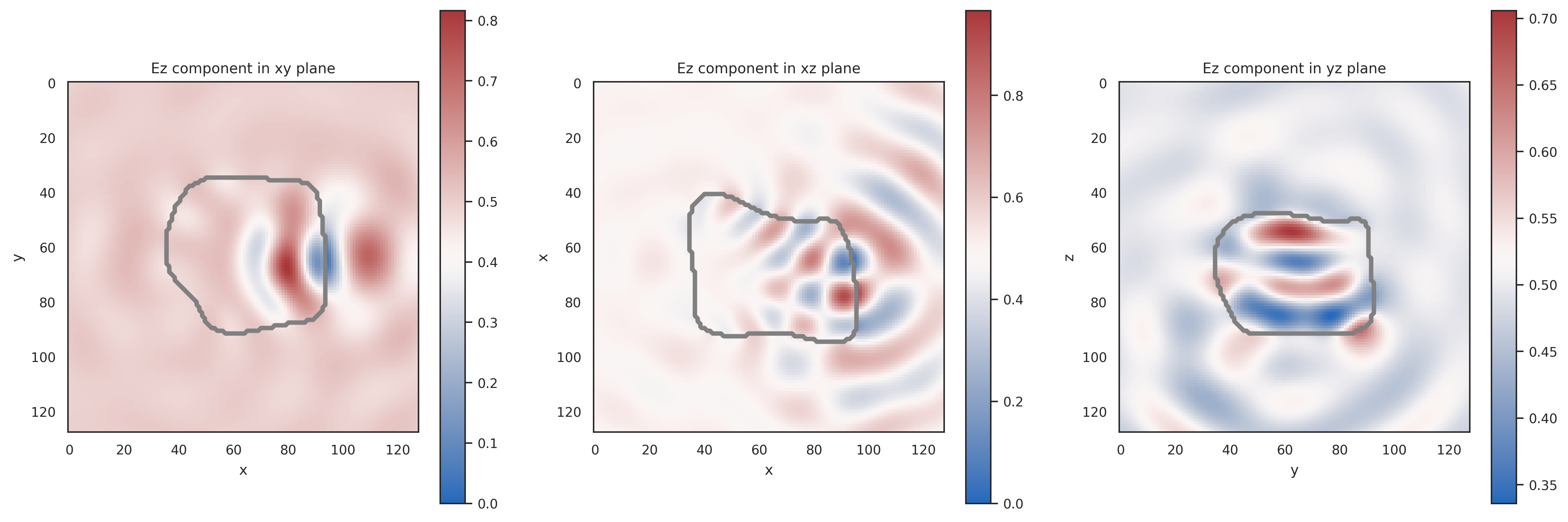}
    \\
    \includegraphics[width=0.9\textwidth]{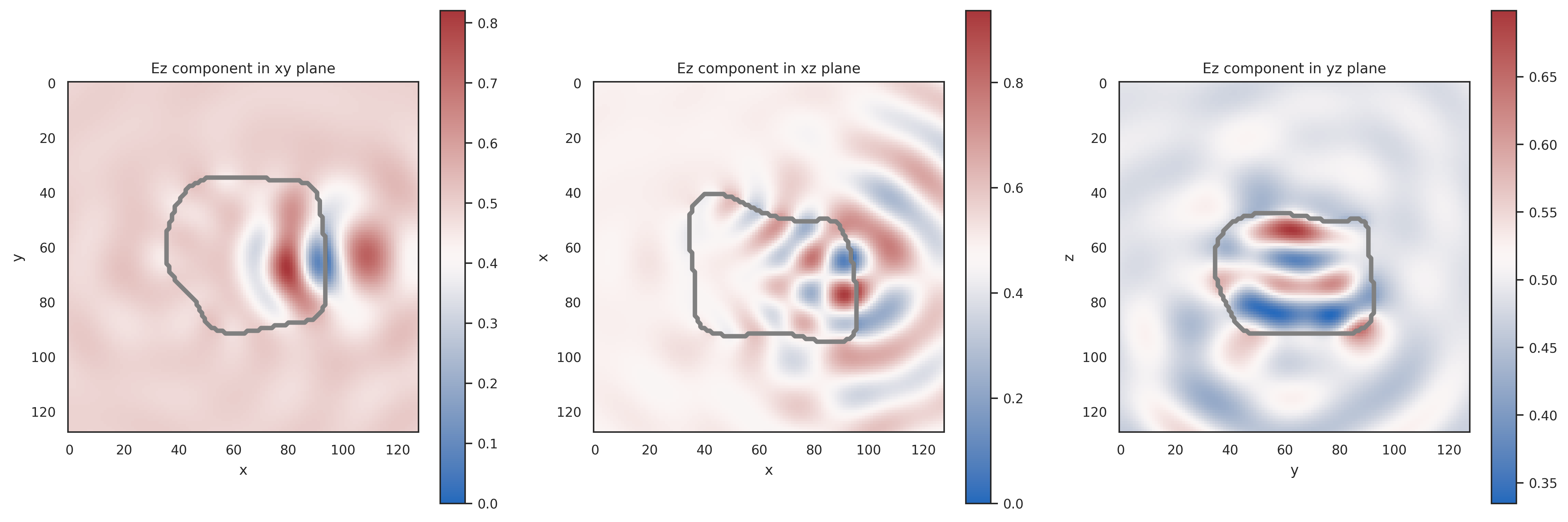}
    \\
    \includegraphics[width=0.9\textwidth]{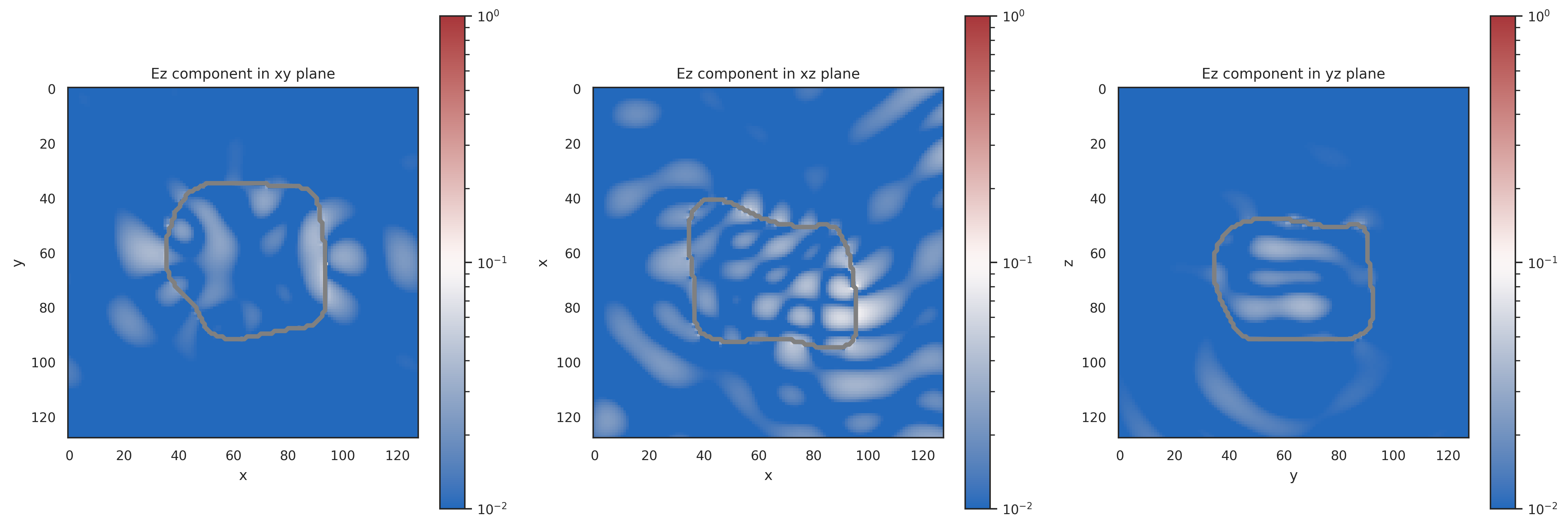}
    \caption{Detailed comparison between the predicted $E_z$ field distributions (top) and the actual, or ground truth, $E_z$ field distributions (middle), illustrating the accuracy of our FDTD solver in simulating electromagnetic wave propagation. The normalized absolute difference is shown in the bottom row.}
    \label{fig:Ez_comparison}
\end{figure}

We validate our solver against the established open source FDTD software Meep \cite{meep} and use the random scatterer dataset by \citet{augensteinNeuralOperatorbasedSurrogate2023a}.
For this experiment, a randomly generated three-dimensional scattering object with refractive index $1.5$ is placed inside a $\SI{6.12}{\micro\meter} \times \SI{6.12}{\micro\meter} \times \SI{6.12}{\micro\meter}$ volume.
The simulation volume has a resolution of $\SI{40}{\nano\meter}$ with the outer 12 voxels being perfectly matched layers.
At the bottom of the simulation volume, a planar wave with wavelength $\SI{1}{\micro\meter}$ is positioned.
The source is active for the the first 10 periods of the simulation and then turned off.
Afterwards, the simulation is run until all the fields are fully decayed, which takes about \SI{100}{\femto\second}.
\Cref{fig:Ez_comparison} shows the central slices of the normalized steady state distribution of the $E_z$ field.
Our solver is able to predict the field distribution to a normalized L1 error of $2.1\%$.

\section{2PP Fabrication Constraints}
\label{sec:2pp}

Though the coupling example introduced by \citet{shenIntegratedMetamaterialsEfficient2014} demonstrates the potential of silicon-based PICs, the fabrication of such devices is time-consuming and expensive.
As an alternative, we consider polymer-based PICs, which can be fabricated using two-photon polymerization (2PP).

\begin{figure}[!t]
    \centering
    \includegraphics[width=0.6\linewidth]{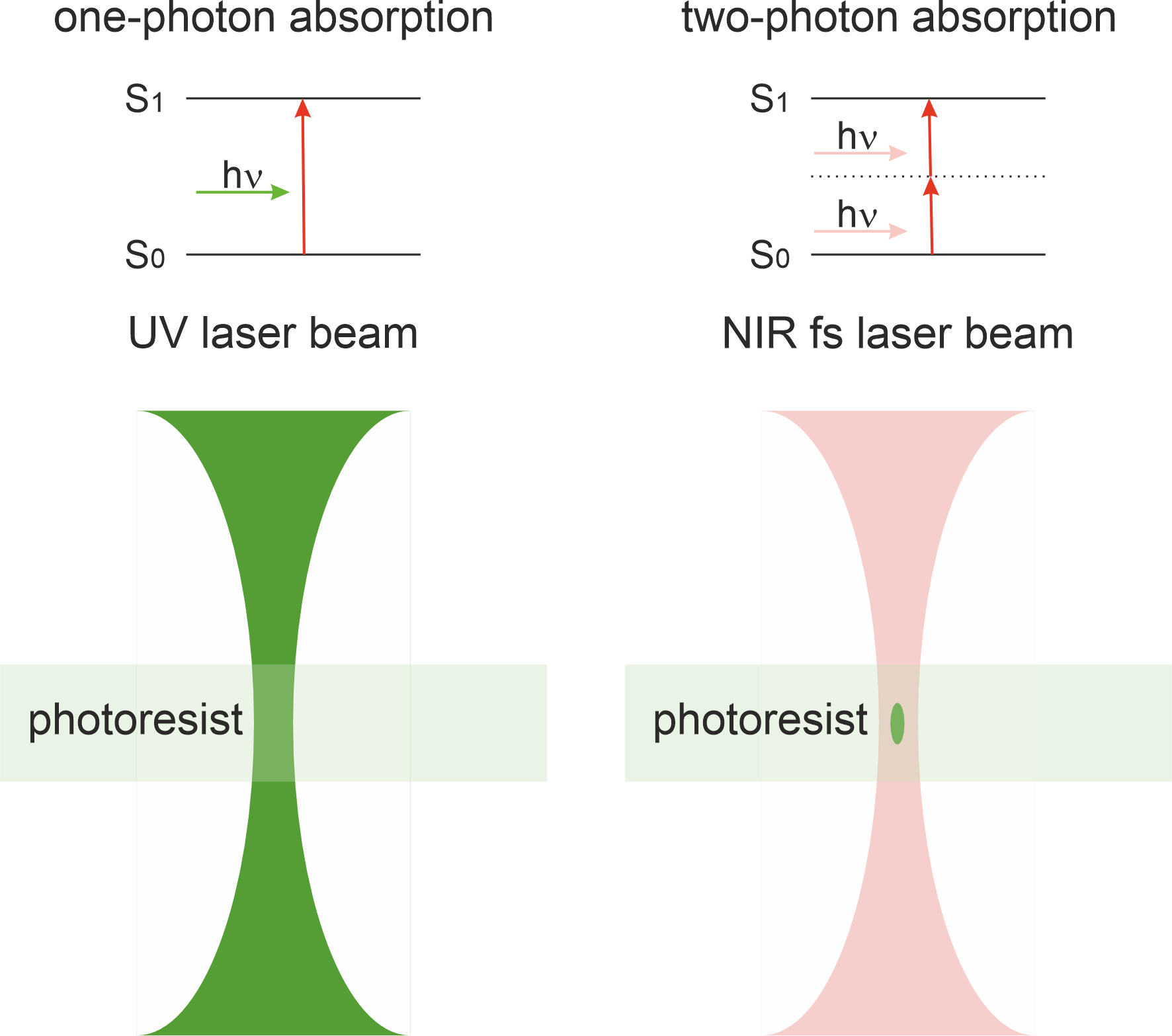}
    \caption{Comparison of conventional laser lithography (left) with 2PP printing (right).}     
    \label{fig:2pp}
\end{figure}

Conventional laser lithography uses a photoresist which absorbs the laser radiation directly.
The absorption of a photon triggers a photo initiator molecule, which starts the cross-linking chain reaction of monomers in the whole cross section volume of laser beam and photoresist, see \cref{fig:2pp} left.
2PP in contrast uses a femtosecond laser with long wavelength for which the photoresist is transparent.
However, the ultrashort pulse of such a laser comes with extremely high peak power.
The optical power level of the laser is carefully adjusted so that in the very center of the laser focus the optical intensity exceeds the non-linear threshold for two-photon absorption which triggers the photo initiator molecules as illustrated in \cref{fig:2pp} right.
Therefore, polymerization takes place only in a tiny volume in the center of the laser focus.
This allows for a free voxel-based 3D printing process inside the volume of the photoresist without the need of support structures and at a resolution far beyond the diffraction limit of the optical system \cite{haske65NmFeature2007}.
The non-linear process can actually involve the simultaneous absorption of multiple photons.
It is thus sometimes called multi-photon polymerization instead of 2PP.

\highlight{
Resulting from this manufacturing technique, multiple constraints regarding the feasibility of designs arises.
Firstly, feasible designs have to adhere to a minimum feature size.
We enforce this constraints by optimizing a discrete voxel grid with cells of side length $\SI{500}{\micro\meter}$.
Post-2PP fabrication requires removal of unpolymerized photoresist through a washing process. 
Consequently, the 3D design must avoid fully enclosed cavities that would prevent complete photoresist removal.
Lastly, another constraint is that all structural elements must maintain physical connectivity to the ground.
This constraint arises due to the three-dimensional fabrication capabilities since two-dimensional designs are always guaranteed to be connected to the ground.
The quantization mapping enforcing these new constraints is discussed in the sections below.
}

\begin{algorithm}[!t]
\caption{\highlight{2.5D Quantization Mapping}}\label{alg:2_5d}
\KwIn{Latent parameters $\theta$, valid permittivities $\mathcal{P}$}
\KwOut{Quantized parameters $\hat{\theta}$}
\smallskip
\Comment{Filter for valid configurations}
$\hat{\mathcal{P}} \gets \emptyset $\;
\For{$p \in \mathcal{P}^{|Z|}$}{
\If{$\nexists i: (p_i = 1 $ and $\exists j > i: p_j \neq 1)$}{
$\hat{\mathcal{P}} \gets \hat{\mathcal{P}} \cup \{p\}$\;
}}
\smallskip
\Comment{Find closest configuration}
\For{$(x,y) \in X \times Y$}{
\For{$p \in \hat{\mathcal{P}}$}{
\smallskip
\Comment{Difference of voxels adjacent in z}
$d_p \gets \sum_{z=1}^{|Z| - 1} |(p_z - p_{z+1}) - (\theta_{x,y,z} - \theta_{x,y,z+1})| $\;
\medskip
\Comment{Difference of average permittivity}
$d_p \gets d_p + \frac{1}{|Z|}|\sum_{i=1}^{|Z|}\theta_{x,y,z} - \sum_{i=1}^{|Z|}p_z|$\;
}
$\hat{\theta}_{x, y} \gets \argmin\limits_{{p} \in \mathcal{P}^{|Z|}} d_p$\;
}
\Return{$\hat{\theta}$}\;
\end{algorithm}

\section{Optimization Details}
\highlight{
Building upon the 2D quantization approach outlined in \cref{alg:2d}, we develop a 2.5D quantization mapping displayed in \cref{alg:2_5d} that enforces a pillar structure.
The process begins by determining all valid configurations, followed by calculating the distance between each pixel and possible configuration.
Our distance metric does not only consider average permittivity, but also accounts for permittivity variations between voxels along the z-axis.
This approach ensures that the quantization accurately captures the structural characteristics represented by the latent parameters.
}

\begin{algorithm}[!t]
\caption{\highlight{3D Quantization Mapping}}\label{alg:3d}
\KwIn{Latent parameters $\theta$, valid permittivities $\mathcal{P}$}
\KwOut{Quantized parameters $\hat{\theta}$}
\For{$(x,y,z) \in X \times Y \times Z$}{
    $\hat{\theta}_{x, y, z} \gets \argmin\limits_{{p} \in \mathcal{P}} |\theta_{x, y, z} - p|$\;
}
\Comment{Remove enclosed cavities}
\While{$\exists \hat{\theta}_{x,y,z} = 1$ not connected to outside}{
    \For{$(x,y,z) \in X \times Y \times Z$}{
        \If{$\hat{\theta}_{x,y,z} = 1$ and not connected to outside}
        {
            $\hat{\theta}_{x,y,z} = p_i$, \,\,$p_i \in \mathcal{P}, p_i \neq 1$\;
        }
    }
}
\Comment{Eliminate structurally unsupported polymer voxels}
\For{$(x,y,z) \in X \times Y \times Z$}{
    \If{$\hat{\theta}_{x,y,z} \neq 1$ and not connected to ground}
    {
        $\hat{\theta}_{x,y,z} = 1$\;
    }
}
\Return{$\hat{\theta}$}\;
\end{algorithm}

\highlight{
In 3D designs, a simple quantization mapping might produce invalid structures where voxels may not be connected to the ground or air cavities within the design are present.
To ensure feasibility, we implement a two-step process.
Firstly, we fill any enclosed air cavities with material and afterwards remove all unsupported structures.
This approach guarantees a valid design since the removal of unsupported structures cannot create new air cavities.
The complete procedure is detailed in \cref{alg:3d}.
}

\begin{figure}[!t]
    \centering
    \includegraphics[width=0.9\linewidth]{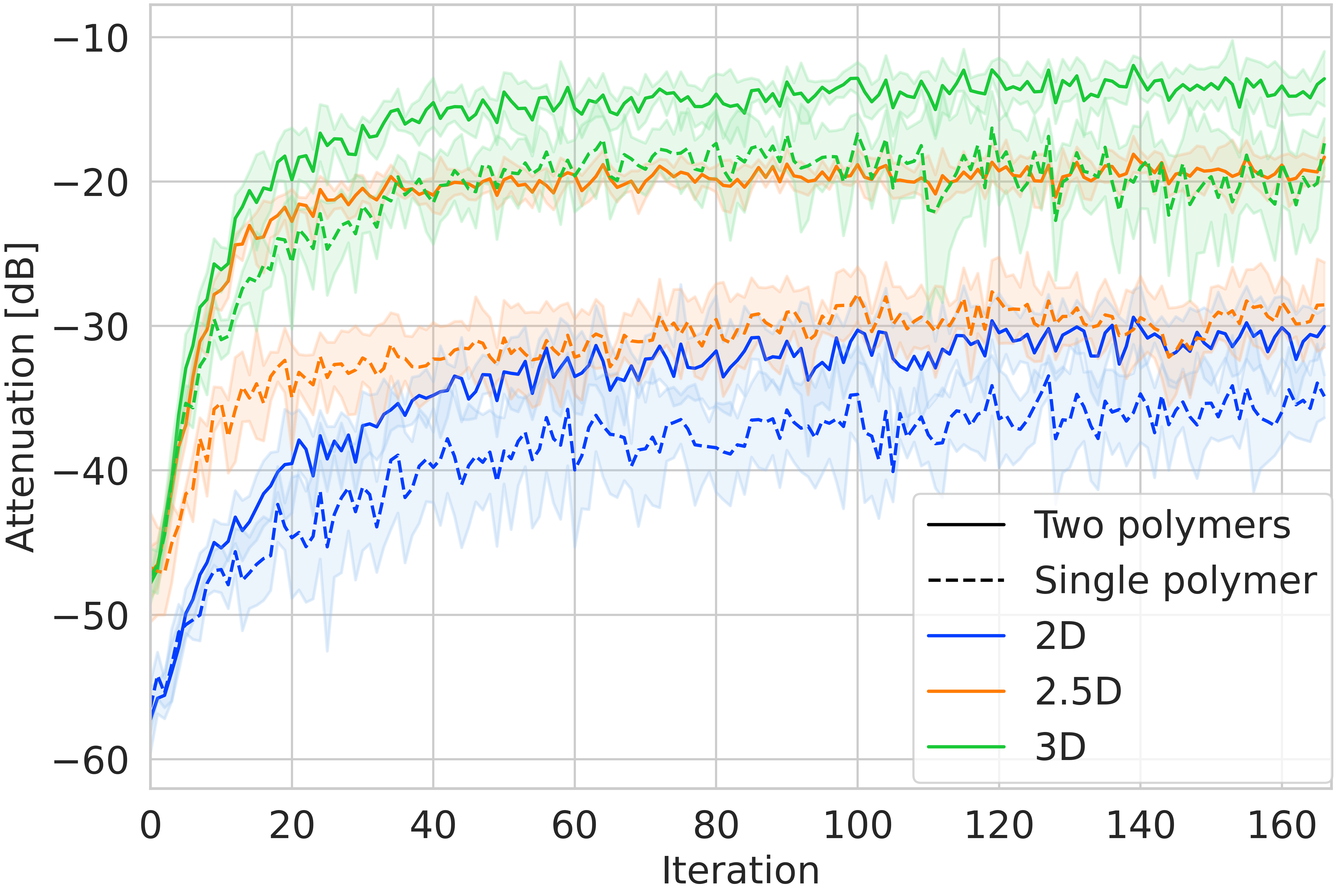}
    \caption{Average attenuation during the optimization of the coupling devices. The shaded area indicates the standard deviation over 5 random seeds and 3 random source calibration offsets per iteration.}     
    \label{fig:optimization}
\end{figure}

To demonstrate the robustness of our stochastic optimization we provide the full optimization curve for 5 seeds per design space in \cref{fig:optimization}.
\highlight{
Each iteration takes around $\SI{110}{\second}$ on a NVIDIA A100 which results in a total wall-clock time of about 15 hours.
The peak memory requirement of this optimization is about 14GB.
Therefore, it would also be feasible to run this optimization on a consumer-grade graphics card like the NVIDIA RTX 4060 TI.
}

\begin{figure}[!t]
    \centering
    \begin{subfigure}{0.45\linewidth}
        \centering
        \includegraphics[width=\linewidth]{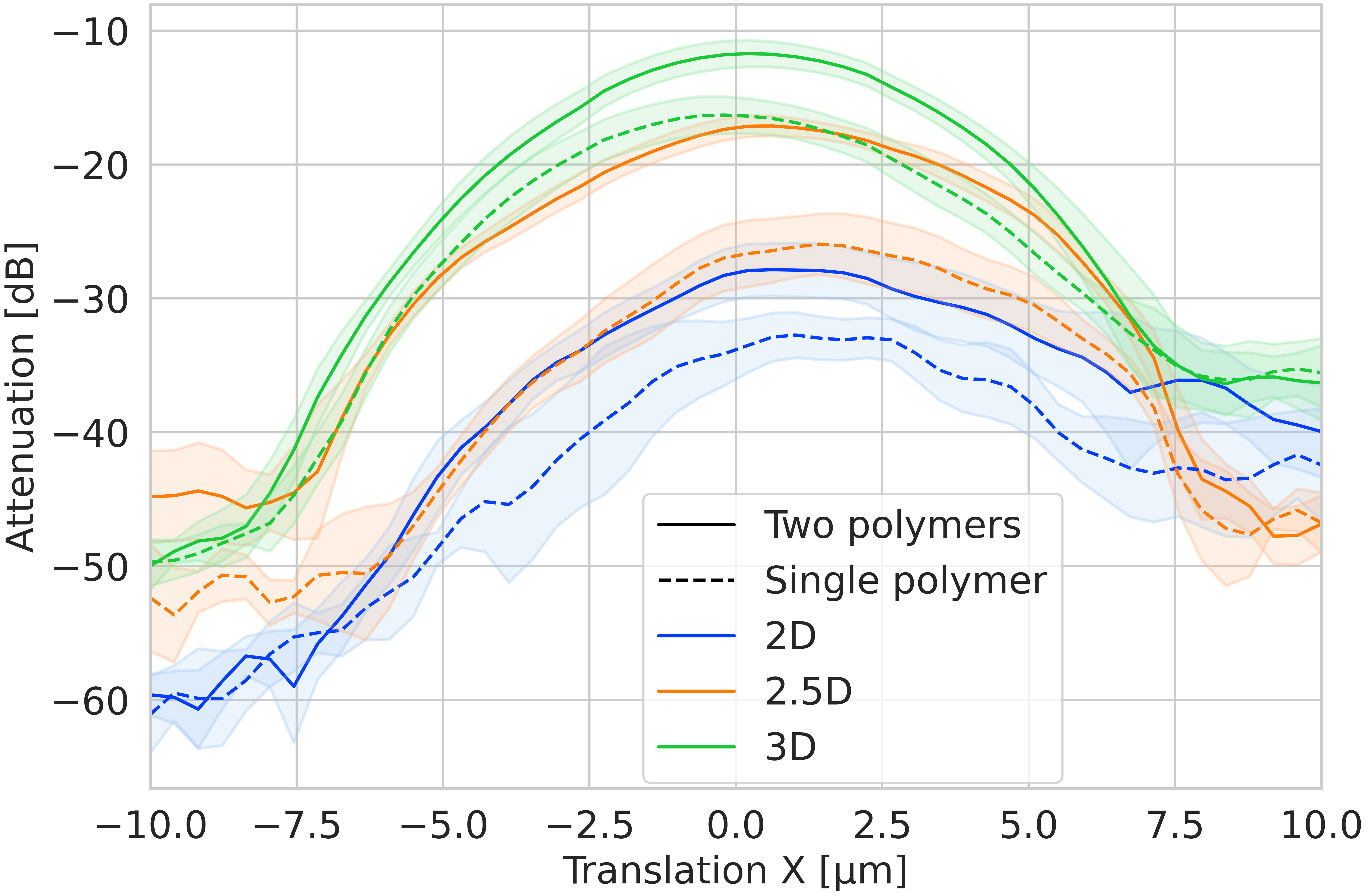}
        \caption{Translation in X-axis}
        \label{fig:trans_x}
    \end{subfigure}
    \begin{subfigure}{0.45\linewidth}
        \centering
        \includegraphics[width=\linewidth]{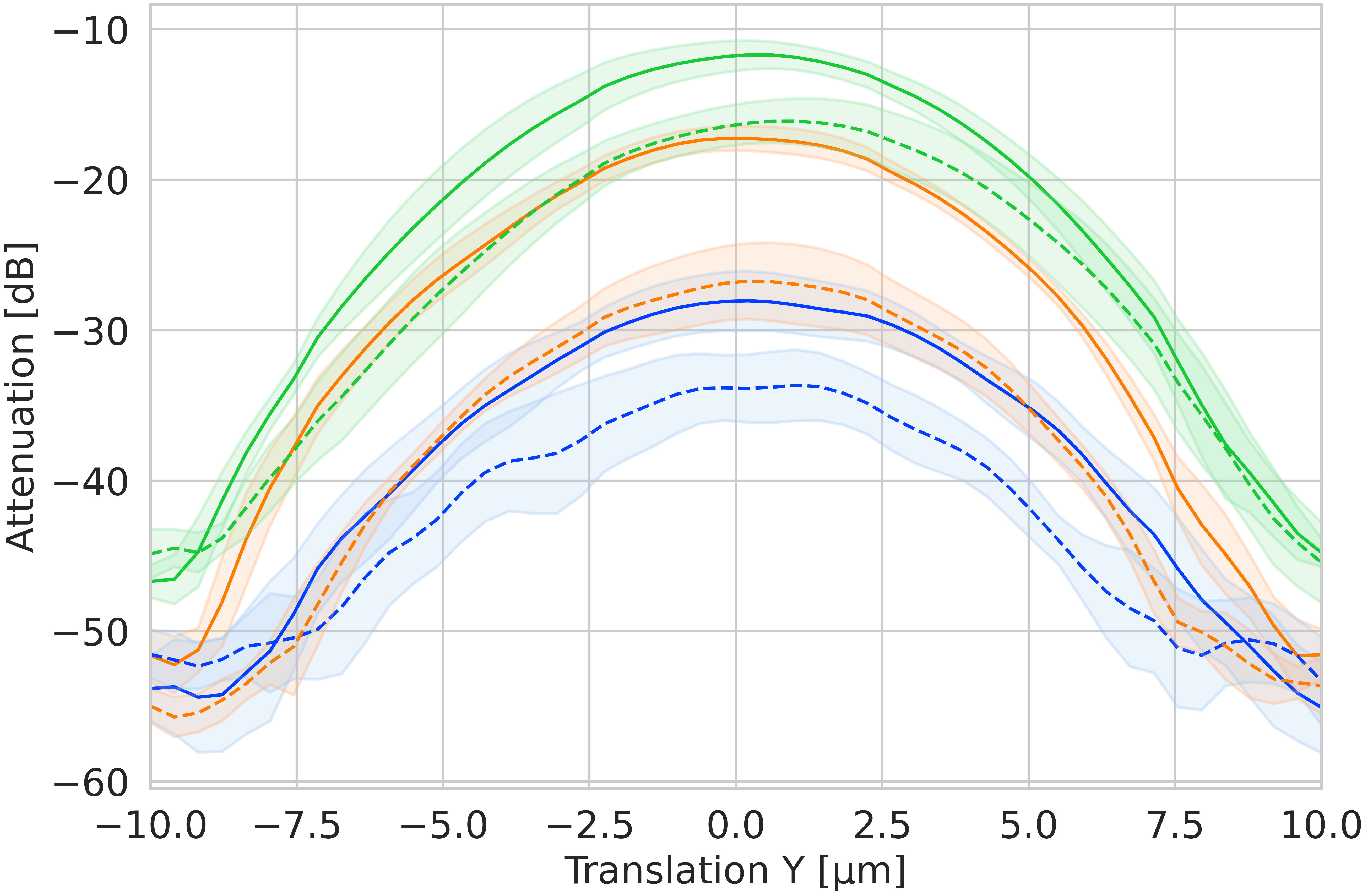}
        \caption{Translation in Y-axis}
        \label{fig:trans_y}
    \end{subfigure}
    \begin{subfigure}{0.45\linewidth}
        \centering
        \includegraphics[width=\linewidth]{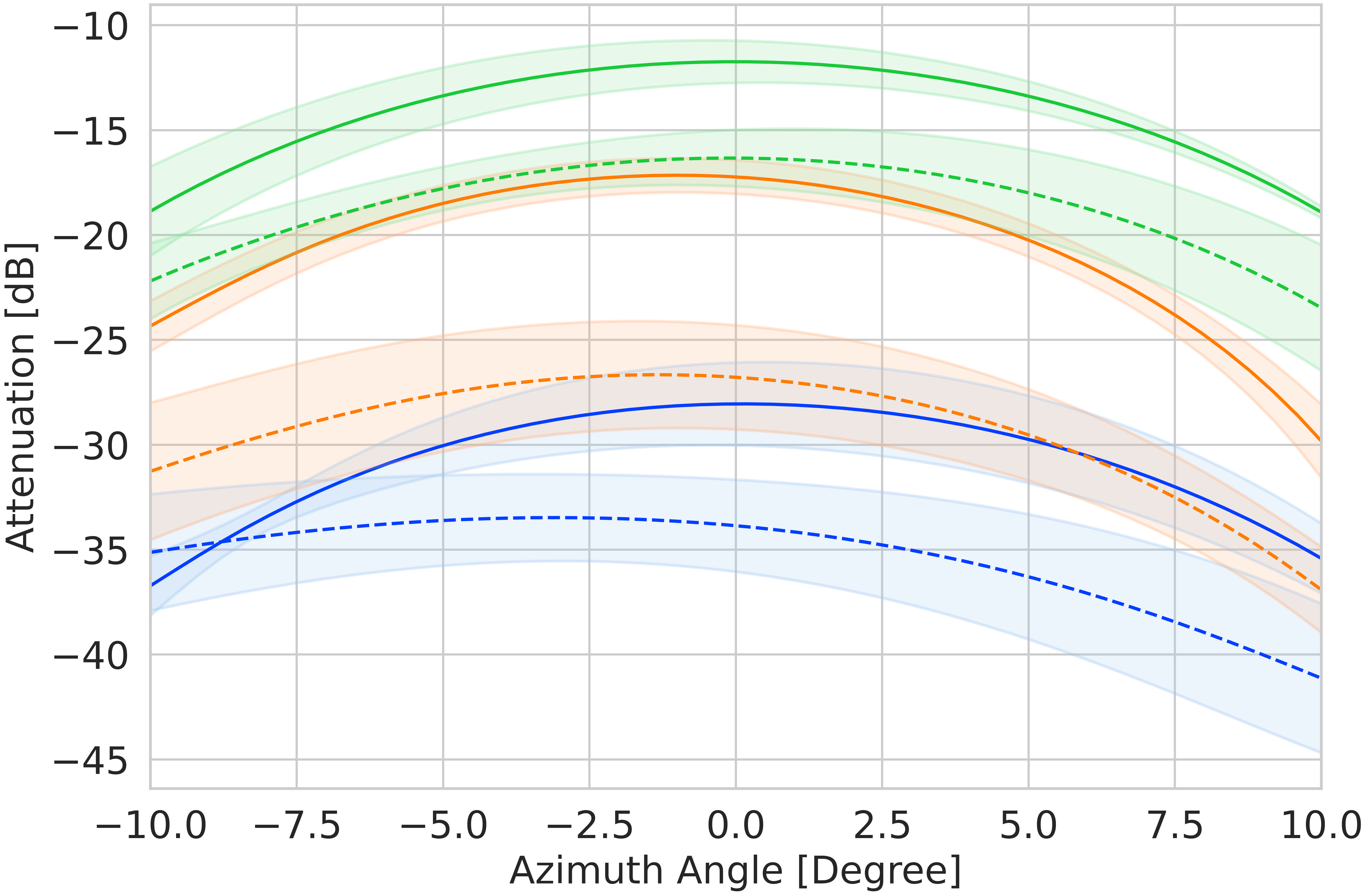}
        \caption{Rotation of Azimuth angle}
        \label{fig:angle_az}
    \end{subfigure}
    \begin{subfigure}{0.45\linewidth}
        \centering
        \includegraphics[width=\linewidth]{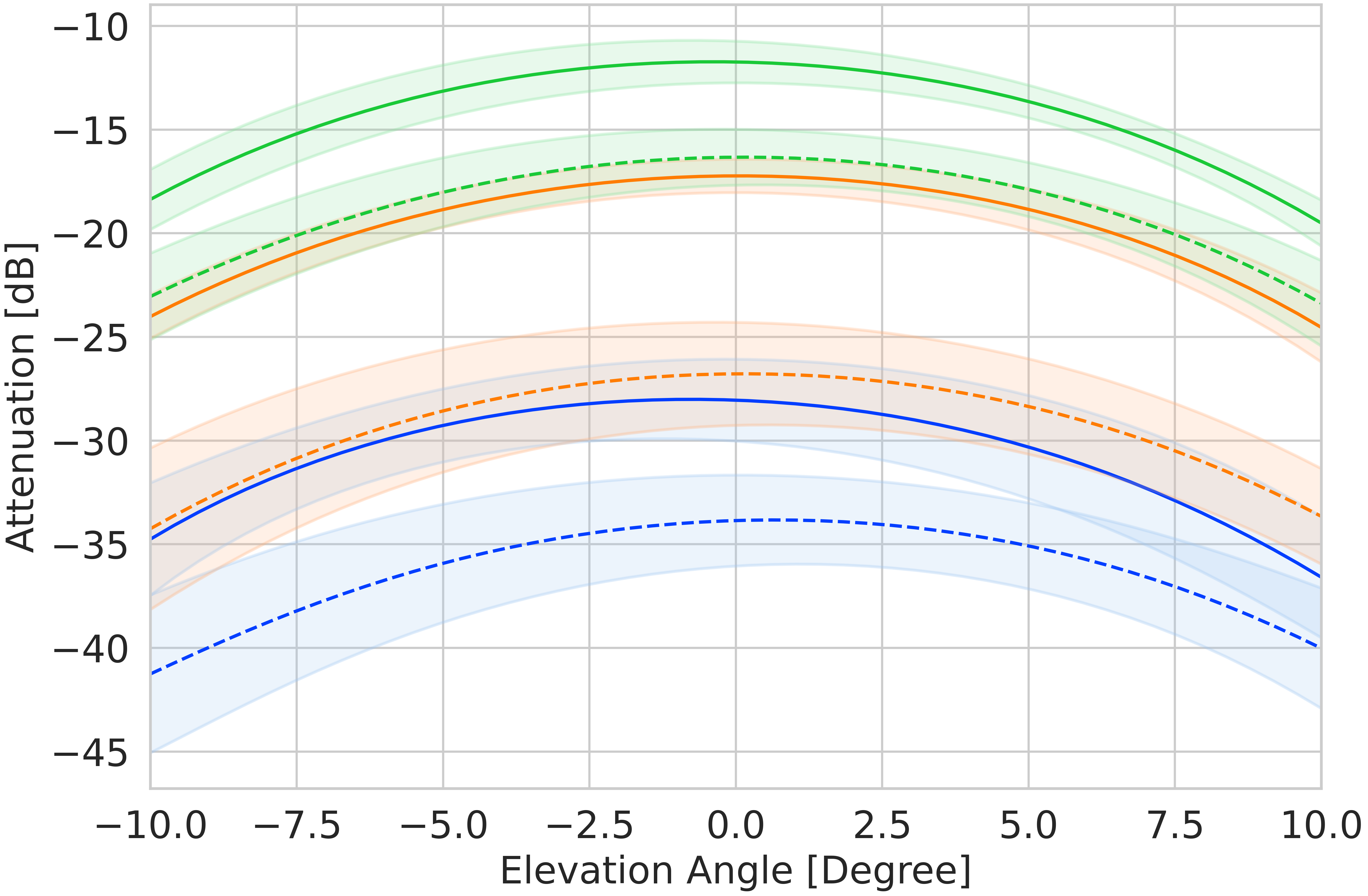}
        \caption{Rotation of Elevation angle}
        \label{fig:angle_el}
    \end{subfigure}
    \begin{subfigure}{0.45\linewidth}
        \centering
        \includegraphics[width=\linewidth]{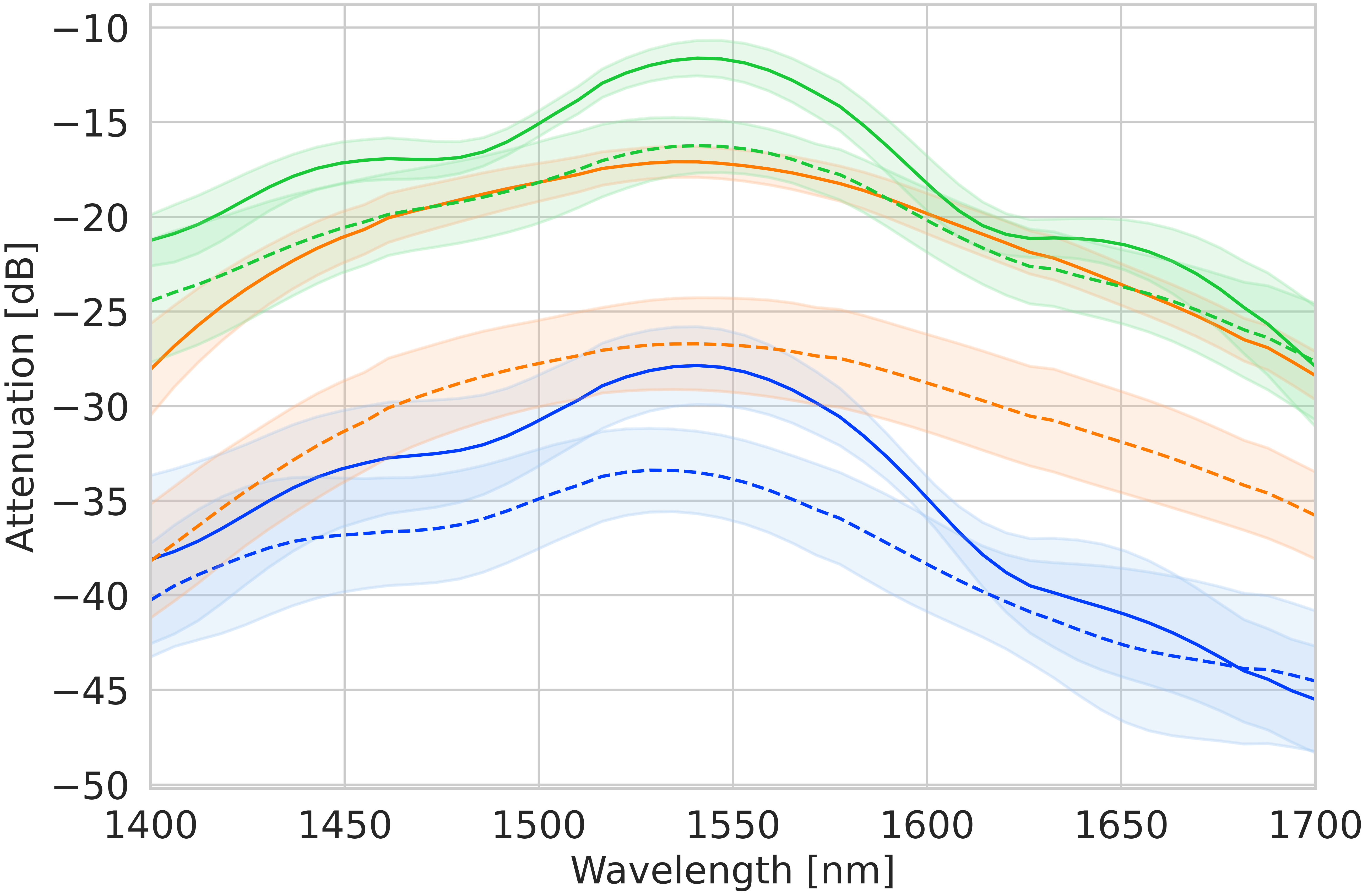}
        \caption{Variation of Wavelength.}
        \label{fig:wavelength}
    \end{subfigure}
    \caption{
    \highlight{
    Analysis of the fully optimized coupling devices in 2D, 2.5D and 3D using either a single or two different polymers variations of the incoming light.
    The source is translated in X- and Y-axis and rotated using the azimuth and elevation angle.
    Additionally, the influence of the source wavelength on the coupling efficiency is evaluated.
    The mean and standard deviation are calculated over five random seeds.
    }
    }
    \label{fig:analysis}
\end{figure}

\highlight{
Using the fully optimized designs, we perform an extensive analysis on the influence of translation and rotational errors in the experimental setup, as well as the distribution under wavelength variations of the source.
The results of this analysis are displayed in \cref{fig:analysis}.
All of the optimized structures are robust to minor changes in translation, rotation and wavelength.
A rotation of 5 degrees, translation of $\SI{2}{\micro\meter}$ or a wavelength variation of $\SI{20}{\nano\meter}$ only result in a drop in efficiency of about $\SI{-2}{\deci\bel}$.
Larger variations also lead to a more drastic reduction in coupling efficiency.
For example, a translation of $\SI{10}{\micro\meter}$ would place only about half of the source above the device, which results in an efficiency drop of $\SI{40}{\deci\bel}$ for the 3D multi-material devices.
In conclusion, our devices are robust to small variations in the experimental setup, but larger variations should be avoided.
}

%% file: quantized_inverse_design.bbl
\providecommand{\latin}[1]{#1}
\makeatletter
\providecommand{\doi}
  {\begingroup\let\do\@makeother\dospecials
  \catcode`\{=1 \catcode`\}=2 \doi@aux}
\providecommand{\doi@aux}[1]{\endgroup\texttt{#1}}
\makeatother
\providecommand*\mcitethebibliography{\thebibliography}
\csname @ifundefined\endcsname{endmcitethebibliography}  {\let\endmcitethebibliography\endthebibliography}{}
\begin{mcitethebibliography}{51}
\providecommand*\natexlab[1]{#1}
\providecommand*\mciteSetBstSublistMode[1]{}
\providecommand*\mciteSetBstMaxWidthForm[2]{}
\providecommand*\mciteBstWouldAddEndPuncttrue
  {\def\EndOfBibitem{\unskip.}}
\providecommand*\mciteBstWouldAddEndPunctfalse
  {\let\EndOfBibitem\relax}
\providecommand*\mciteSetBstMidEndSepPunct[3]{}
\providecommand*\mciteSetBstSublistLabelBeginEnd[3]{}
\providecommand*\EndOfBibitem{}
\mciteSetBstSublistMode{f}
\mciteSetBstMaxWidthForm{subitem}{(\alph{mcitesubitemcount})}
\mciteSetBstSublistLabelBeginEnd
  {\mcitemaxwidthsubitemform\space}
  {\relax}
  {\relax}

\bibitem[Jandura \latin{et~al.}(2017)Jandura, Gaso, and Pudis]{janduraPolymerBasedDevices2017}
Jandura,~D.; Gaso,~P.; Pudis,~D. Polymer {{Based Devices}} for {{Photonics}} on the {{Chip}}. \emph{Communications - Scientific letters of the University of Zilina} \textbf{2017}, \emph{19}, 16--20\relax
\mciteBstWouldAddEndPuncttrue
\mciteSetBstMidEndSepPunct{\mcitedefaultmidpunct}
{\mcitedefaultendpunct}{\mcitedefaultseppunct}\relax
\EndOfBibitem
\bibitem[Gonzalez-Hernandez \latin{et~al.}(2023)Gonzalez-Hernandez, Varapnickas, Bertoncini, Liberale, and Malinauskas]{gonzalez-hernandezMicroOptics3D2023}
Gonzalez-Hernandez,~D.; Varapnickas,~S.; Bertoncini,~A.; Liberale,~C.; Malinauskas,~M. Micro-{{Optics 3D Printed}} via {{Multi}}-{{Photon Laser Lithography}}. \emph{Advanced Optical Materials} \textbf{2023}, \emph{11}, 2201701\relax
\mciteBstWouldAddEndPuncttrue
\mciteSetBstMidEndSepPunct{\mcitedefaultmidpunct}
{\mcitedefaultendpunct}{\mcitedefaultseppunct}\relax
\EndOfBibitem
\bibitem[Zhang \latin{et~al.}(2013)Zhang, Hosseini, Lin, Subbaraman, and Chen]{zhangPolymerBasedHybridIntegratedPhotonic2013}
Zhang,~X.; Hosseini,~A.; Lin,~X.; Subbaraman,~H.; Chen,~R.~T. Polymer-{{Based Hybrid-Integrated Photonic Devices}} for {{Silicon On-Chip Modulation}} and {{Board-Level Optical Interconnects}}. \emph{IEEE Journal of Selected Topics in Quantum Electronics} \textbf{2013}, \emph{19}, 196--210\relax
\mciteBstWouldAddEndPuncttrue
\mciteSetBstMidEndSepPunct{\mcitedefaultmidpunct}
{\mcitedefaultendpunct}{\mcitedefaultseppunct}\relax
\EndOfBibitem
\bibitem[Swatowski(2017)]{swatowskiPOLYMERWAVEGUIDEMANUFACTURING2017}
Swatowski,~B. Polymer {{Waveguide Manufacturing}} and {{Printed Circuit Board Integration}}. Master of {{Science}} in {{Electrical Engineering}}, Michigan Technological University, Houghton, Michigan, 2017\relax
\mciteBstWouldAddEndPuncttrue
\mciteSetBstMidEndSepPunct{\mcitedefaultmidpunct}
{\mcitedefaultendpunct}{\mcitedefaultseppunct}\relax
\EndOfBibitem
\bibitem[G{\"u}nther(2017)]{guenther_2017}
G{\"u}nther,~A. Optische {{Strukturen}} Zur {{Einkopplung}} Und {{Lichtverteilung}} in Vollst{\"a}ndig Polymerbasierten, Planaren {{Systemen}}. Ph.D.\ thesis, Gottfried Wilhelm Leibniz Universit{\"a}t Hannover, Hannover, 2017\relax
\mciteBstWouldAddEndPuncttrue
\mciteSetBstMidEndSepPunct{\mcitedefaultmidpunct}
{\mcitedefaultendpunct}{\mcitedefaultseppunct}\relax
\EndOfBibitem
\bibitem[Marchetti \latin{et~al.}(2019)Marchetti, Lacava, Carroll, Gradkowski, and Minzioni]{marchettiCouplingStrategiesSilicon2019}
Marchetti,~R.; Lacava,~C.; Carroll,~L.; Gradkowski,~K.; Minzioni,~P. Coupling Strategies for Silicon Photonics Integrated Chips [{{Invited}}]. \emph{Photonics Research} \textbf{2019}, \emph{7}, 201\relax
\mciteBstWouldAddEndPuncttrue
\mciteSetBstMidEndSepPunct{\mcitedefaultmidpunct}
{\mcitedefaultendpunct}{\mcitedefaultseppunct}\relax
\EndOfBibitem
\bibitem[Son \latin{et~al.}(2018)Son, Han, Park, Kwon, and Yu]{sonHighefficiencyBroadbandLight2018}
Son,~G.; Han,~S.; Park,~J.; Kwon,~K.; Yu,~K. High-Efficiency Broadband Light Coupling between Optical Fibers and Photonic Integrated Circuits. \emph{Nanophotonics} \textbf{2018}, \emph{7}, 1845--1864\relax
\mciteBstWouldAddEndPuncttrue
\mciteSetBstMidEndSepPunct{\mcitedefaultmidpunct}
{\mcitedefaultendpunct}{\mcitedefaultseppunct}\relax
\EndOfBibitem
\bibitem[{Kane Yee}(1966)]{kaneyeeNumericalSolutionInitial1966}
{Kane Yee} Numerical Solution of Initial Boundary Value Problems Involving Maxwell's Equations in Isotropic Media. \emph{IEEE Transactions on Antennas and Propagation} \textbf{1966}, \emph{14}, 302--307\relax
\mciteBstWouldAddEndPuncttrue
\mciteSetBstMidEndSepPunct{\mcitedefaultmidpunct}
{\mcitedefaultendpunct}{\mcitedefaultseppunct}\relax
\EndOfBibitem
\bibitem[Yang \latin{et~al.}(2021)Yang, Mayer, Bunz, Blasco, and Wegener]{yangMultimaterialMultiphoton3D2021}
Yang,~L.; Mayer,~F.; Bunz,~U. H.~F.; Blasco,~E.; Wegener,~M. Multi-Material Multi-Photon {{3D}} Laser Micro- and Nanoprinting. \emph{Light: Advanced Manufacturing} \textbf{2021}, \emph{2}, 1\relax
\mciteBstWouldAddEndPuncttrue
\mciteSetBstMidEndSepPunct{\mcitedefaultmidpunct}
{\mcitedefaultendpunct}{\mcitedefaultseppunct}\relax
\EndOfBibitem
\bibitem[{Gonzalez-Hernandez} \latin{et~al.}(2023){Gonzalez-Hernandez}, {Sanchez-Padilla}, Gailevicius, Thodika, Juodkazis, Brasselet, and Malinauskas]{gonzalez-hernandezSingleStep3D2023}
{Gonzalez-Hernandez},~D.; {Sanchez-Padilla},~B.; Gailevicius,~D.; Thodika,~S.~C.; Juodkazis,~S.; Brasselet,~E.; Malinauskas,~M. Single-{{Step 3D Printing}} of {{Micro}}-{{Optics}} with {{Adjustable Refractive Index}} by {{Ultrafast Laser Nanolithography}}. \textbf{2023}, \relax
\mciteBstWouldAddEndPunctfalse
\mciteSetBstMidEndSepPunct{\mcitedefaultmidpunct}
{}{\mcitedefaultseppunct}\relax
\EndOfBibitem
\bibitem[Yu \latin{et~al.}(2023)Yu, Prudnikau, Lesnyak, and Kirchner]{yuQuantumDotsFacilitate2023}
Yu,~Y.; Prudnikau,~A.; Lesnyak,~V.; Kirchner,~R. Quantum {{Dots Facilitate 3D Two}}-{{Photon Laser Lithography}}. \emph{Advanced Materials} \textbf{2023}, \emph{35}, 2211702\relax
\mciteBstWouldAddEndPuncttrue
\mciteSetBstMidEndSepPunct{\mcitedefaultmidpunct}
{\mcitedefaultendpunct}{\mcitedefaultseppunct}\relax
\EndOfBibitem
\bibitem[{Lumerical Solution Inc.}()]{AnsysLumericalFDTD}
{Lumerical Solution Inc.} FDTD Solution, A Commercial Professional Software. https://www.lumerical.com\relax
\mciteBstWouldAddEndPuncttrue
\mciteSetBstMidEndSepPunct{\mcitedefaultmidpunct}
{\mcitedefaultendpunct}{\mcitedefaultseppunct}\relax
\EndOfBibitem
\bibitem[Oskooi \latin{et~al.}(2010)Oskooi, Roundy, Ibanescu, Bermel, Joannopoulos, and Johnson]{meep}
Oskooi,~A.~F.; Roundy,~D.; Ibanescu,~M.; Bermel,~P.; Joannopoulos,~J.; Johnson,~S.~G. Meep: A flexible free-software package for electromagnetic simulations by the FDTD method. \emph{Computer Physics Communications} \textbf{2010}, \emph{181}, 687--702\relax
\mciteBstWouldAddEndPuncttrue
\mciteSetBstMidEndSepPunct{\mcitedefaultmidpunct}
{\mcitedefaultendpunct}{\mcitedefaultseppunct}\relax
\EndOfBibitem
\bibitem[Agrawal \latin{et~al.}(2019)Agrawal, Modi, Passos, Lavoie, Agarwal, Shankar, Ganichev, Levenberg, Hong, Monga, and Cai]{agrawalTensorFlowEagerMultiStage2019}
Agrawal,~A.; Modi,~A.~N.; Passos,~A.; Lavoie,~A.; Agarwal,~A.; Shankar,~A.; Ganichev,~I.; Levenberg,~J.; Hong,~M.; Monga,~R.; Cai,~S. {{TensorFlow Eager}}: {{A Multi-Stage}}, {{Python-Embedded DSL}} for {{Machine Learning}}. 2019\relax
\mciteBstWouldAddEndPuncttrue
\mciteSetBstMidEndSepPunct{\mcitedefaultmidpunct}
{\mcitedefaultendpunct}{\mcitedefaultseppunct}\relax
\EndOfBibitem
\bibitem[Paszke \latin{et~al.}(2019)Paszke, Gross, Massa, Lerer, Bradbury, Chanan, Killeen, Lin, Gimelshein, Antiga, Desmaison, K{\"o}pf, Yang, DeVito, Raison, Tejani, Chilamkurthy, Steiner, Fang, Bai, and Chintala]{paszkePyTorchImperativeStyle2019}
Paszke,~A. \latin{et~al.}  {{PyTorch}}: {{An Imperative Style}}, {{High-Performance Deep Learning Library}}. 2019\relax
\mciteBstWouldAddEndPuncttrue
\mciteSetBstMidEndSepPunct{\mcitedefaultmidpunct}
{\mcitedefaultendpunct}{\mcitedefaultseppunct}\relax
\EndOfBibitem
\bibitem[Bradbury \latin{et~al.}(2018)Bradbury, Frostig, Hawkins, Johnson, Leary, Maclaurin, Necula, Paszke, VanderPlas, {Wanderman-Milne}, and Zhang]{jax2018github}
Bradbury,~J.; Frostig,~R.; Hawkins,~P.; Johnson,~M.~J.; Leary,~C.; Maclaurin,~D.; Necula,~G.; Paszke,~A.; VanderPlas,~J.; {Wanderman-Milne},~S.; Zhang,~Q. {{JAX}}: Composable Transformations of {{Python}}+{{NumPy}} Programs. 2018\relax
\mciteBstWouldAddEndPuncttrue
\mciteSetBstMidEndSepPunct{\mcitedefaultmidpunct}
{\mcitedefaultendpunct}{\mcitedefaultseppunct}\relax
\EndOfBibitem
\bibitem[Schubert \latin{et~al.}(2022)Schubert, Cheung, Williamson, Spyra, and Alexander]{schubertInverseDesignPhotonic2022}
Schubert,~M.~F.; Cheung,~A. K.~C.; Williamson,~I. A.~D.; Spyra,~A.; Alexander,~D.~H. Inverse {{Design}} of {{Photonic Devices}} with {{Strict Foundry Fabrication Constraints}}. \emph{ACS Photonics} \textbf{2022}, \emph{9}, 2327--2336\relax
\mciteBstWouldAddEndPuncttrue
\mciteSetBstMidEndSepPunct{\mcitedefaultmidpunct}
{\mcitedefaultendpunct}{\mcitedefaultseppunct}\relax
\EndOfBibitem
\bibitem[Maxwell(1865)]{maxwellVIIIDynamicalTheory1865}
Maxwell,~J.~C. {{VIII}}. {{A}} Dynamical Theory of the Electromagnetic Field. \emph{Philosophical Transactions of the Royal Society of London} \textbf{1865}, \emph{155}, 459--512\relax
\mciteBstWouldAddEndPuncttrue
\mciteSetBstMidEndSepPunct{\mcitedefaultmidpunct}
{\mcitedefaultendpunct}{\mcitedefaultseppunct}\relax
\EndOfBibitem
\bibitem[Courant \latin{et~al.}(1928)Courant, Friedrichs, and Lewy]{courantUberPartiellenDifferenzengleichungen1928}
Courant,~R.; Friedrichs,~K.; Lewy,~H. {\"U}ber Die Partiellen {{Differenzengleichungen}} Der Mathematischen {{Physik}}. \emph{Mathematische Annalen} \textbf{1928}, \emph{100}, 32--74\relax
\mciteBstWouldAddEndPuncttrue
\mciteSetBstMidEndSepPunct{\mcitedefaultmidpunct}
{\mcitedefaultendpunct}{\mcitedefaultseppunct}\relax
\EndOfBibitem
\bibitem[Roden and Gedney(2000)Roden, and Gedney]{rodenConvolutionPMLCPML2000}
Roden,~J.~A.; Gedney,~S.~D. Convolution {{PML}} ({{CPML}}): {{An}} Efficient {{FDTD}} Implementation of the {{CFS-PML}} for Arbitrary Media. \emph{Microwave and Optical Technology Letters} \textbf{2000}, \emph{27}, 334--339\relax
\mciteBstWouldAddEndPuncttrue
\mciteSetBstMidEndSepPunct{\mcitedefaultmidpunct}
{\mcitedefaultendpunct}{\mcitedefaultseppunct}\relax
\EndOfBibitem
\bibitem[Taflove and Hagness(1998)Taflove, and Hagness]{tafloveComputationalElectrodynamicsFiniteDifference1998}
Taflove,~A.; Hagness,~S.~C. \emph{Computational {{Electrodynamics}} - {{The Finite-Difference Time-Domain Method}}}; 1998\relax
\mciteBstWouldAddEndPuncttrue
\mciteSetBstMidEndSepPunct{\mcitedefaultmidpunct}
{\mcitedefaultendpunct}{\mcitedefaultseppunct}\relax
\EndOfBibitem
\bibitem[Michaels()]{EMoptDocumentationEMopt}
Michaels,~A. Emopt: Electromagnetic optimization toolbox. https://github.com/anstmichaels/emopt\relax
\mciteBstWouldAddEndPuncttrue
\mciteSetBstMidEndSepPunct{\mcitedefaultmidpunct}
{\mcitedefaultendpunct}{\mcitedefaultseppunct}\relax
\EndOfBibitem
\bibitem[{Lalau-Keraly} \latin{et~al.}(2013){Lalau-Keraly}, Bhargava, Miller, and Yablonovitch]{lalau-keralyAdjointShapeOptimization2013}
{Lalau-Keraly},~C.~M.; Bhargava,~S.; Miller,~O.~D.; Yablonovitch,~E. Adjoint Shape Optimization Applied to Electromagnetic Design. \emph{Optics Express} \textbf{2013}, \emph{21}, 21693\relax
\mciteBstWouldAddEndPuncttrue
\mciteSetBstMidEndSepPunct{\mcitedefaultmidpunct}
{\mcitedefaultendpunct}{\mcitedefaultseppunct}\relax
\EndOfBibitem
\bibitem[Christiansen and Sigmund(2021)Christiansen, and Sigmund]{christiansenInverseDesignPhotonics2021}
Christiansen,~R.~E.; Sigmund,~O. Inverse Design in Photonics by Topology Optimization: Tutorial. \emph{Journal of the Optical Society of America B} \textbf{2021}, \emph{38}, 496\relax
\mciteBstWouldAddEndPuncttrue
\mciteSetBstMidEndSepPunct{\mcitedefaultmidpunct}
{\mcitedefaultendpunct}{\mcitedefaultseppunct}\relax
\EndOfBibitem
\bibitem[{Guoqiang Shen} \latin{et~al.}(2003){Guoqiang Shen}, Tam, Nikolova, and Bakr]{guoqiangshenAdjointSensitivityTechnique2003}
{Guoqiang Shen}; Tam,~H.; Nikolova,~N.; Bakr,~M. Adjoint Sensitivity Technique for {{FDTD}} Methods on Structured Grids. {{IEEE Antennas}} and {{Propagation Society International Symposium}}. {{Digest}}. {{Held}} in Conjunction with: {{USNC}}/{{CNC}}/{{URSI North American Radio Sci}}. {{Meeting}} ({{Cat}}. {{No}}.{{03CH37450}}). Columbus, OH, USA, 2003; pp 746--749\relax
\mciteBstWouldAddEndPuncttrue
\mciteSetBstMidEndSepPunct{\mcitedefaultmidpunct}
{\mcitedefaultendpunct}{\mcitedefaultseppunct}\relax
\EndOfBibitem
\bibitem[Bakr and Nikolova(2003)Bakr, and Nikolova]{bakrAdjointVariableMethod2003}
Bakr,~M.; Nikolova,~N. An Adjoint Variable Method for Frequency Domain {{TLM}} Problems with Conducting Boundaries. \emph{IEEE Microwave and Wireless Components Letters} \textbf{2003}, \emph{13}, 408--410\relax
\mciteBstWouldAddEndPuncttrue
\mciteSetBstMidEndSepPunct{\mcitedefaultmidpunct}
{\mcitedefaultendpunct}{\mcitedefaultseppunct}\relax
\EndOfBibitem
\bibitem[Nikolova \latin{et~al.}(2004)Nikolova, Tam, and Bakr]{nikolovaSensitivityAnalysisFDTD2004}
Nikolova,~N.; Tam,~H.; Bakr,~M. Sensitivity {{Analysis With}} the {{FDTD Method}} on {{Structured Grids}}. \emph{IEEE Transactions on Microwave Theory and Techniques} \textbf{2004}, \emph{52}, 1207--1216\relax
\mciteBstWouldAddEndPuncttrue
\mciteSetBstMidEndSepPunct{\mcitedefaultmidpunct}
{\mcitedefaultendpunct}{\mcitedefaultseppunct}\relax
\EndOfBibitem
\bibitem[Nikolova \latin{et~al.}(2006)Nikolova, {Ying Li}, {Yan Li}, and Bakr]{nikolovaSensitivityAnalysisScattering2006}
Nikolova,~N.; {Ying Li}; {Yan Li}; Bakr,~M. Sensitivity Analysis of Scattering Parameters with Electromagnetic Time-Domain Simulators. \emph{IEEE Transactions on Microwave Theory and Techniques} \textbf{2006}, \emph{54}, 1598--1610\relax
\mciteBstWouldAddEndPuncttrue
\mciteSetBstMidEndSepPunct{\mcitedefaultmidpunct}
{\mcitedefaultendpunct}{\mcitedefaultseppunct}\relax
\EndOfBibitem
\bibitem[Swillam \latin{et~al.}(2007)Swillam, Bakr, Nikolova, and Li]{swillamAdjointSensitivityAnalysis2007}
Swillam,~M.~A.; Bakr,~M.~H.; Nikolova,~N.~K.; Li,~X. Adjoint {{Sensitivity Analysis}} of {{Dielectric Discontinuities Using FDTD}}. \emph{Electromagnetics} \textbf{2007}, \emph{27}, 123--140\relax
\mciteBstWouldAddEndPuncttrue
\mciteSetBstMidEndSepPunct{\mcitedefaultmidpunct}
{\mcitedefaultendpunct}{\mcitedefaultseppunct}\relax
\EndOfBibitem
\bibitem[Yasuda \latin{et~al.}(2019)Yasuda, Yamada, and Nishiwaki]{yasudaDesignMethodSpatiotemporal2019}
Yasuda,~H.; Yamada,~T.; Nishiwaki,~S. A Design Method of Spatiotemporal Optical Pulse Using Level-set Based Time Domain Topology Optimization. \emph{International Journal for Numerical Methods in Engineering} \textbf{2019}, \emph{117}, 605--622\relax
\mciteBstWouldAddEndPuncttrue
\mciteSetBstMidEndSepPunct{\mcitedefaultmidpunct}
{\mcitedefaultendpunct}{\mcitedefaultseppunct}\relax
\EndOfBibitem
\bibitem[Tang \latin{et~al.}(2023)Tang, Lim, Ossiander, Yin, and Capasso]{tangTimeReversalDifferentiation2023}
Tang,~R.~J.; Lim,~S. W.~D.; Ossiander,~M.; Yin,~X.; Capasso,~F. Time {{Reversal Differentiation}} of {{FDTD}} for {{Photonic Inverse Design}}. \emph{ACS Photonics} \textbf{2023}, \emph{10}, 4140--4150\relax
\mciteBstWouldAddEndPuncttrue
\mciteSetBstMidEndSepPunct{\mcitedefaultmidpunct}
{\mcitedefaultendpunct}{\mcitedefaultseppunct}\relax
\EndOfBibitem
\bibitem[Luce \latin{et~al.}(2023)Luce, Alaee, Knorr, and Marquardt]{luceMergingAutomaticDifferentiation2023}
Luce,~A.; Alaee,~R.; Knorr,~F.; Marquardt,~F. Merging Automatic Differentiation and the Adjoint Method for Photonic Inverse Design. 2023\relax
\mciteBstWouldAddEndPuncttrue
\mciteSetBstMidEndSepPunct{\mcitedefaultmidpunct}
{\mcitedefaultendpunct}{\mcitedefaultseppunct}\relax
\EndOfBibitem
\bibitem[Hooten \latin{et~al.}(2023)Hooten, Sun, Gantz, Fiorentino, Beausoleil, and Van~Vaerenbergh]{hootenAutomaticDifferentiationAccelerated2023}
Hooten,~S.; Sun,~P.; Gantz,~L.; Fiorentino,~M.; Beausoleil,~R.~G.; Van~Vaerenbergh,~T. Automatic Differentiation Accelerated Shape Optimization Approaches to Photonic Inverse Design on Rectilinear Simulation Grids. 2023\relax
\mciteBstWouldAddEndPuncttrue
\mciteSetBstMidEndSepPunct{\mcitedefaultmidpunct}
{\mcitedefaultendpunct}{\mcitedefaultseppunct}\relax
\EndOfBibitem
\bibitem[Ballew \latin{et~al.}(2023)Ballew, Roberts, Zheng, and Faraon]{ballewConstrainingContinuousTopology2023}
Ballew,~C.; Roberts,~G.; Zheng,~T.; Faraon,~A. Constraining {{Continuous Topology Optimizations}} to {{Discrete Solutions}} for {{Photonic Applications}}. \emph{ACS Photonics} \textbf{2023}, acsphotonics.2c00862\relax
\mciteBstWouldAddEndPuncttrue
\mciteSetBstMidEndSepPunct{\mcitedefaultmidpunct}
{\mcitedefaultendpunct}{\mcitedefaultseppunct}\relax
\EndOfBibitem
\bibitem[Schubert \latin{et~al.}(2022)Schubert, Cheung, Williamson, Spyra, and Alexander]{schubertInverseDesignPhotonic2022a}
Schubert,~M.~F.; Cheung,~A. K.~C.; Williamson,~I. A.~D.; Spyra,~A.; Alexander,~D.~H. Inverse {{Design}} of {{Photonic Devices}} with {{Strict Foundry Fabrication Constraints}}. \emph{ACS Photonics} \textbf{2022}, \emph{9}, 2327--2336\relax
\mciteBstWouldAddEndPuncttrue
\mciteSetBstMidEndSepPunct{\mcitedefaultmidpunct}
{\mcitedefaultendpunct}{\mcitedefaultseppunct}\relax
\EndOfBibitem
\bibitem[Paolini \latin{et~al.}(2022)Paolini, De~Marinis, Cococcioni, Valcarenghi, Maggiani, and Andriolli]{paoliniPhotonicawareNeuralNetworks2022}
Paolini,~E.; De~Marinis,~L.; Cococcioni,~M.; Valcarenghi,~L.; Maggiani,~L.; Andriolli,~N. Photonic-Aware Neural Networks. \emph{Neural Computing \& Applications} \textbf{2022}, \emph{34}, 15589--15601\relax
\mciteBstWouldAddEndPuncttrue
\mciteSetBstMidEndSepPunct{\mcitedefaultmidpunct}
{\mcitedefaultendpunct}{\mcitedefaultseppunct}\relax
\EndOfBibitem
\bibitem[Bengio \latin{et~al.}(2013)Bengio, L{\'e}onard, and Courville]{bengioEstimatingPropagatingGradients2013}
Bengio,~Y.; L{\'e}onard,~N.; Courville,~A. Estimating or {{Propagating Gradients Through Stochastic Neurons}} for {{Conditional Computation}}. 2013\relax
\mciteBstWouldAddEndPuncttrue
\mciteSetBstMidEndSepPunct{\mcitedefaultmidpunct}
{\mcitedefaultendpunct}{\mcitedefaultseppunct}\relax
\EndOfBibitem
\bibitem[Yin \latin{et~al.}(2019)Yin, Lyu, Zhang, Osher, Qi, and Xin]{yinUnderstandingStraightEstimatorTraining2019}
Yin,~P.; Lyu,~J.; Zhang,~S.; Osher,~S.; Qi,~Y.; Xin,~J. Understanding {{Straight-Through Estimator}} in {{Training Activation Quantized Neural Nets}}. 2019\relax
\mciteBstWouldAddEndPuncttrue
\mciteSetBstMidEndSepPunct{\mcitedefaultmidpunct}
{\mcitedefaultendpunct}{\mcitedefaultseppunct}\relax
\EndOfBibitem
\bibitem[Schoenbauer \latin{et~al.}(2024)Schoenbauer, Moro, Lew, and Howard]{Schoenbauer2024CustomGE}
Schoenbauer,~M.; Moro,~D.; Lew,~L.; Howard,~A. Custom Gradient Estimators are Straight-Through Estimators in Disguise. \emph{ArXiv} \textbf{2024}, \emph{abs/2405.05171}\relax
\mciteBstWouldAddEndPuncttrue
\mciteSetBstMidEndSepPunct{\mcitedefaultmidpunct}
{\mcitedefaultendpunct}{\mcitedefaultseppunct}\relax
\EndOfBibitem
\bibitem[Kingma and Ba(2015)Kingma, and Ba]{KingmaB14}
Kingma,~D.~P.; Ba,~J. Adam: {A} Method for Stochastic Optimization. 3rd International Conference on Learning Representations, {ICLR} 2015, San Diego, CA, USA, May 7-9, 2015, Conference Track Proceedings. 2015\relax
\mciteBstWouldAddEndPuncttrue
\mciteSetBstMidEndSepPunct{\mcitedefaultmidpunct}
{\mcitedefaultendpunct}{\mcitedefaultseppunct}\relax
\EndOfBibitem
\bibitem[{micro resist technology GmbH}()]{microresisttechnologyProcessingGuidelines}
{micro resist technology GmbH} ma-N 400 and ma-N 1400 - Negative Tone Photoresists. \url{https://www.microresist.com}\relax
\mciteBstWouldAddEndPuncttrue
\mciteSetBstMidEndSepPunct{\mcitedefaultmidpunct}
{\mcitedefaultendpunct}{\mcitedefaultseppunct}\relax
\EndOfBibitem
\bibitem[Nesterov(1983)]{1370862715914709505}
Nesterov,~Y. A method for solving the convex programming problem with convergence rate o(1/k2). 1983\relax
\mciteBstWouldAddEndPuncttrue
\mciteSetBstMidEndSepPunct{\mcitedefaultmidpunct}
{\mcitedefaultendpunct}{\mcitedefaultseppunct}\relax
\EndOfBibitem
\bibitem[Loshchilov and Hutter(2017)Loshchilov, and Hutter]{LoshchilovH17}
Loshchilov,~I.; Hutter,~F. {SGDR:} Stochastic Gradient Descent with Warm Restarts. 5th International Conference on Learning Representations, {ICLR} 2017, Toulon, France, April 24-26, 2017, Conference Track Proceedings. 2017\relax
\mciteBstWouldAddEndPuncttrue
\mciteSetBstMidEndSepPunct{\mcitedefaultmidpunct}
{\mcitedefaultendpunct}{\mcitedefaultseppunct}\relax
\EndOfBibitem
\bibitem[Sorrentino \latin{et~al.}(1993)Sorrentino, Roselli, and Mezzanotte]{sorrentinoTimeReversalFinite1993}
Sorrentino,~R.; Roselli,~L.; Mezzanotte,~P. Time Reversal in Finite Difference Time Domain Method. \emph{IEEE Microwave and Guided Wave Letters} \textbf{1993}, \emph{3}, 402--404\relax
\mciteBstWouldAddEndPuncttrue
\mciteSetBstMidEndSepPunct{\mcitedefaultmidpunct}
{\mcitedefaultendpunct}{\mcitedefaultseppunct}\relax
\EndOfBibitem
\bibitem[Shen \latin{et~al.}(2014)Shen, Wang, Polson, and Menon]{shenIntegratedMetamaterialsEfficient2014}
Shen,~B.; Wang,~P.; Polson,~R.; Menon,~R. Integrated Metamaterials for Efficient and Compact Free-Space-to-Waveguide Coupling. \emph{Optics Express} \textbf{2014}, \emph{22}, 27175\relax
\mciteBstWouldAddEndPuncttrue
\mciteSetBstMidEndSepPunct{\mcitedefaultmidpunct}
{\mcitedefaultendpunct}{\mcitedefaultseppunct}\relax
\EndOfBibitem
\bibitem[Lee and Chang(1995)Lee, and Chang]{leeApplicationComputationalGeometry1995}
Lee,~Y.-S.; Chang,~T.-C. Application of computational geometry in optimizing 2.{5D} and {3D} {NC} surface machining. \emph{Computers in Industry} \textbf{1995}, \emph{26}, 41--59\relax
\mciteBstWouldAddEndPuncttrue
\mciteSetBstMidEndSepPunct{\mcitedefaultmidpunct}
{\mcitedefaultendpunct}{\mcitedefaultseppunct}\relax
\EndOfBibitem
\bibitem[Siu \latin{et~al.}(2019)Siu, Kim, Miele, and Follmer]{siuShapeCADAccessible3D2019}
Siu,~A.~F.; Kim,~S.; Miele,~J.~A.; Follmer,~S. {shapeCAD}: {An} {Accessible} {3D} {Modelling} {Workflow} for the {Blind} and {Visually}-{Impaired} {Via} 2.{5D} {Shape} {Displays}. The 21st {International} {ACM} {SIGACCESS} {Conference} on {Computers} and {Accessibility}. Pittsburgh PA USA, 2019; pp 342--354\relax
\mciteBstWouldAddEndPuncttrue
\mciteSetBstMidEndSepPunct{\mcitedefaultmidpunct}
{\mcitedefaultendpunct}{\mcitedefaultseppunct}\relax
\EndOfBibitem
\bibitem[Siu \latin{et~al.}(2018)Siu, Miele, and Follmer]{siuAccessibleCADWorkflow2018}
Siu,~A.~F.; Miele,~J.; Follmer,~S. An {Accessible} {CAD} {Workflow} {Using} {Programming} of {3D} {Models} and {Preview} {Rendering} in {A} 2.{5D} {Shape} {Display}. Proceedings of the 20th {International} {ACM} {SIGACCESS} {Conference} on {Computers} and {Accessibility}. Galway Ireland, 2018; pp 343--345\relax
\mciteBstWouldAddEndPuncttrue
\mciteSetBstMidEndSepPunct{\mcitedefaultmidpunct}
{\mcitedefaultendpunct}{\mcitedefaultseppunct}\relax
\EndOfBibitem
\bibitem[Augenstein \latin{et~al.}(2023)Augenstein, Rep{\"a}n, and Rockstuhl]{augensteinNeuralOperatorbasedSurrogate2023a}
Augenstein,~Y.; Rep{\"a}n,~T.; Rockstuhl,~C. A Neural Operator-Based Surrogate Solver for Free-Form Electromagnetic Inverse Design. \emph{ACS Photonics} \textbf{2023}, \emph{10}, 1547--1557\relax
\mciteBstWouldAddEndPuncttrue
\mciteSetBstMidEndSepPunct{\mcitedefaultmidpunct}
{\mcitedefaultendpunct}{\mcitedefaultseppunct}\relax
\EndOfBibitem
\bibitem[Haske \latin{et~al.}(2007)Haske, Chen, Hales, Dong, Barlow, Marder, and Perry]{haske65NmFeature2007}
Haske,~W.; Chen,~V.~W.; Hales,~J.~M.; Dong,~W.; Barlow,~S.; Marder,~S.~R.; Perry,~J.~W. 65 Nm Feature Sizes Using Visible Wavelength 3-{{D}} Multiphoton Lithography. \emph{Optics Express} \textbf{2007}, \emph{15}, 3426\relax
\mciteBstWouldAddEndPuncttrue
\mciteSetBstMidEndSepPunct{\mcitedefaultmidpunct}
{\mcitedefaultendpunct}{\mcitedefaultseppunct}\relax
\EndOfBibitem
\end{mcitethebibliography}
